\documentclass[aps,prd,twocolumn,showpacs,nofootinbib,amsmath,amssymb,amsfonts,showkeys]{revtex4}

\usepackage{graphicx}
\usepackage{bm}

\begin{document}

\title{Non-linear power spectra of dark and luminous \\
  matter in halo model of structure formation}

\author{Yurij Kulinich}
 \email{kul@astro.franko.lviv.ua}
\author{Bohdan Novosyadlyj}
 \email{novos@astro.franko.lviv.ua}
\author{Stepan Apunevych}
 \email{apus@astro.franko.lviv.ua}
\affiliation{Astronomical Observatory of Ivan Franko National University of Lviv, Kyryla i Methodia str., 8, Lviv, 79005, Ukraine}

\date{\today}

\begin{abstract}
  The late stages of large-scale structure evolution are treated
  semi-analytically within the framework of modified halo model. We
  suggest simple yet accurate approximation for relating the
  non-linear amplitude to linear one for spherical density
  perturbation. For halo concentration parameter, $c$, a new computation
  technique is proposed, which eliminates the need of interim
  evaluation of the $z_{col}$. Validity of the technique is proved for
   $\Lambda$CDM and $\Lambda$WDM cosmologies. Also, the
  parameters for Sheth-Tormen mass function are estimated. The modified and extended halo
  model is applied for determination of non-linear power spectrum of
  dark matter, as well as for galaxy power spectrum estimation. The
  semi-analytical techniques for dark matter power spectrum are
  verified by comparison with data from numerical simulations. Also, 
  the predictions for the galaxy power spectra are confronted with
  'observed' data from PSCz and SDSS galaxy catalogs, good accordance
  is found.
\end{abstract}
\maketitle

\section{Introduction}

A commonly accepted inflationary paradigm states that the large-scale
structure (LSS) of the Universe is formed through evolution of density
perturbations driven by gravitational instability. 
At some moment the growth of small-scale perturbations switches to
non-linear regime. The treatment of linear regime is quite simple, the
linear power spectrum (transfer function) for $k<0.1$ h/Mpc 
can be readily computed with percent accuracy for any
feasible cosmology. However, it is not so for the smaller scales, due
to the non-linear terms in equations and complexity of physics of
baryonic component (hydrodynamics, radiation transfer, thermal and
chemical evolution). %
This paper is aimed for the development of technique capable to build
a bridge between the initial (linear) matter power spectrum and
observable (inherently non-linear) galaxy power spectrum. The
treatment is based on halo model complemented by analytical
approximations, the results are tested and verified against data of
N-body simulations.

Within the scenario commonly referred as ``standard'' (see
\cite{b71}), the gravitational potential of collisionless dark matter
inhomogeneities governs the baryonic matter until the baryonic matter
power spectrum reaches the dark matter's one in amplitude. At some
moment, a non-linear perturbation with amplitude exceeding some
critical one detaches from background expansion, reaches a turnaround
point and starts to collapse due to self-gravity. Subsequently the
violent relaxation takes place, which brings the system into the
virial equilibrium, so the halo of dark matter is formed. Then, the
baryonic gas starts to cool down, followed by clumping into clouds and
ignition of the luminous tracers within halos (see \cite{b4,b6} for
details).

Thus, the spatial distribution of luminous matter should strongly
correlate with one of halos. The correlation is confirmed by large
simulations, which take into account the baryon physics and particle
dynamics of dark matter \cite{b57,b58}, and by semi-analytic models of
galaxy formation \cite{b59,b60,b61} as well.  The numerical techniques
require the considerable computing power, whereas the purely
analytical are found unreliable and inaccurate. Thus, the ``hybrid''
approach seems to be optimal, combining the analytical model of galaxy
formation \cite{b62,b63} with dark matter ``merger trees'' extracted
from simulations.
Another way is to extract the halo and subhalo statistics from
simulations for comparison with galaxy populations in large galaxy survey. 
Such techniques are based on
conditional luminosity function (CLF) \cite{b19,b32,b33}, conditional
mass function (CMF) \cite{b20} and stellar mass to halo mass relation
\cite{b1,b20}.

The halo model is a cornerstone of modern theory of structure
formation. It has been proven to be well-motivated, comprehensive and
provides plausible explanation for observational data and results of
cosmological simulations. It is valid for wide range of cosmologies,
as long as the statistics of primordial density perturbations is
Gaussian. It encompasses the non-linear stage of evolution of density
perturbations as well as the dynamical relaxation processes assuming
that the whole mass is associated with gravitationally bound virialized
halo.

It has shown in papers \cite{b7,b70,b65} that the dark matter
non-linear power spectrum can be evaluated given halo statistics,
their internal structure and spatial distribution. Also
\textit{vice versa}, the initial power spectrum can be reconstructed
by applying the halo model to the data of N-body simulations. As
another example, the halo occupation function, $p(N|M)$, probability
of finding $N$ galaxies within a halo of mass $M$, was used in
\cite{b67,r03} along with halo model to calculate the non-linear galaxy
power spectrum.  However, regardless of overall success of the halo
model\footnote{The details of halo model can be found in \cite{b0}.}
in description of dark matter and galaxy clustering, it is still not
to be considered as complete. For instance, only the latest
enhancements proposed in \cite{b29,r04} take into account the internal
structure of halo as well as the halo shapes \cite{r05,r06}.

To apply the halo model, the \textit{a priori} knowledge 
of the evolution of inhomogeneities from initial state through collapse
to the formation of virialized halo is required. In section II we analyze such
evolution using the spherical perturbation model in order to analytically
relate the amplitude of non-linear spherical density perturbation to
the one of the linear.  Also, the new technique is proposed therein
for computation of concentration parameter $c$ for halo with
Navarro-Frenk-White density profile. In section III the halo mass
function is applied to the dark matter clustering.  In section IV the
galaxy power spectra are estimated and compared to 'observable' ones,
as derived in \cite{b26,b27} from PSCz and SDSS catalogs.
The conclusions are presented in section V.  The computations were
performed for $\Lambda$CDM (cold dark matter with $\Lambda$-term) and
$\Lambda$WDM (warm dark matter with $\Lambda$-term) cosmological
models, some bulk mathematical derivations are separated in
appendices.

\section{Formation of individual spherical halo}

\subsection{Spherical overdensity with arbitrary profile}


In the framework of Tolman's approach \cite{b78}, the spherically symmetrical
inhomogeneity is treated in synchronous gauge (\textit{i.e.} with
regard to the frame comoving to the dust-like matter component), with
space-time interval
\begin{eqnarray}
 ds^2 &=& dt^2 - \frac{y^2(t,R)}{\sqrt{1-KR^2}}dR^2 \nonumber\\ &&-
  x^2(t,R)R^2\left(d\theta^2+\sin^2\theta d\varphi^2\right), \label{Tolman_ds2}
\end{eqnarray}
where $t$ is a proper time of an observer located at $R$, $K$ is space curvature of the Universe as the whole.
 For homogeneous Friedmann Universe $x(t)=y(t)=a(t)$.

The mean matter density, \textit{i.e.} average value over the sphere
of radius $R$, is denoted by $\rho_M(\tau,R)\equiv
\rho_m^0x^{-3}(\tau,R)$, and the matter density at some specific
distance $R$ from the center of perturbation is
$\rho(\tau,R)\equiv\rho_m^0x^{-2}(\tau,R)y^{-1}(\tau,R)$, see \cite{b17,b17a}
for details. Let us define the amplitude of density perturbation as
follows:
\begin{equation}
 \delta_{\rho}(t,R) = \frac{\rho(t,R)-\overline{\rho}(t)}{\overline{\rho}(t)} = \frac{a^3(t)}{x^2(t,R)y(t,R)} - 1,\label{delta_rho}
\end{equation}
the amplitude of the mass perturbation is
\begin{equation}
 \delta_{M}(t,R) = \frac{\rho_M(t,R)-\overline{\rho}(t)}{\overline{\rho}(t)} = \frac{a^3(t)}{x^3(t,R)} - 1.\label{delta_M}
\end{equation}
These amplitudes (see \cite{b17}) are related by
\begin{equation}
 \delta_M(t,R) = \frac{3}{r^{3}(t,R)}\int_0^R \delta_{\rho}(t,R)r^2(t,R)r'(t,R)dR \label{relation1}
\end{equation}
with $r(t,R)=x(t,R)R$, $(')\equiv d/dR$.

For small perturbations the following approximations are valid for
 (\ref{delta_rho}) and (\ref{delta_M}):
\begin{eqnarray}
  \delta_{\rho}(t,R) &\simeq & \delta(t,R) \equiv A(R)D\left(a(t)\right)\label{delta1} \ll 1 ,\\ 
  \delta_M(t,R) &\simeq &\overline{\delta}(t,R)\equiv \frac35\frac{\Omega_K-\Omega_f(R)}{\Omega_m}\, D\left(a(t)\right)\label{delta2} \ll 1 ,
\end{eqnarray}
were $D(a(\tau))$ is the growth factor of linear matter density perturbations
\cite{b12,b18}, defined by
\begin{equation}
D(a) = \frac52 \Omega_ma^{-1}X^{1/2}(a)\int_0^aX^{-3/2}(\tilde{a})d\tilde{a},
  \label{Dl}
\end{equation}
where $X(a) \equiv \Omega_{\Lambda}a^2 + \Omega_ma^{-1} + \Omega_{K}$.
The parameter of local curvature, $\Omega_f(R)$, can be related due to
Eq.\,(\ref{relation1}) with the density profile, $A(R)$, from
(\ref{delta2}) as
\begin{equation}
  \Omega_{f}(R) = -5\Omega_mR^{-3}\int\limits_0^R A(\tilde{R})\tilde{R}^2d\tilde{R} + \Omega_K.
 \label{curvature}
\end{equation}
Thus, either $A(R)$ or $\Omega_f(R)$ should be specified to define
the initial profile of perturbation.

The Einstein's equations for spherical overdensity of dust-like matter in the model with cosmological constant,
$\mathcal{G}^1_1 =\mathcal{G}^2_2 =\Lambda$, yield the
equations for $x(\tau,R)$ and $y(\tau,R)$:
\begin{eqnarray}
  \ddot x & = & \frac32\Omega_{\Lambda}x - \frac12 \frac{\dot x^2}{x} + \frac{1}{x}\frac{\Omega_f}{2},\label{Per1}\\
  \ddot y & = & \frac32\Omega_{\Lambda}y - \left(\frac{\dot x\dot y}{x} - \frac12 \frac{\dot x^2y}{x^2}\right) + \left(\Omega_f + 
    R \frac{\Omega_f'}{2}\right)\frac{1}{x} - \frac{y}{x^2}\frac{\Omega_f}{2} \nonumber \\ 
  && \label{Per2} 
\end{eqnarray}
The overdot denotes a derivative with respect to $\tau=H_0t$. The
first integration of (\ref{Per1}) yields
\begin{equation}
  \dot x^2 -  \frac{\Omega_m}{x} - \Omega_{\Lambda} x^2 = \Omega_f.
  \label{xscale}
\end{equation}
Thus, the amplitude of mass perturbation, (\ref{delta_M}), can be
calculated by integrating over the time just this single equation.

The development of spherical inhomogeneity  can be divided into two 
stages: 1) the expansion stage ($\dot x > 0$, $\dot \rho_M < 0$),
linear and weakly non-linear regime; 2) the collapse stage ($\dot x <
0$, $\dot \rho_M > 0$), entirely non-linear regime. They are separated
by the moment of turnaround, when $\dot x = 0$ ($\dot \rho_M =0$).

The evolution of perturbation at non-linear stage is convenient to be
treated by confronting with the evolution of some fictitious linear
perturbation extrapolated beyond the linear stage. Therefore, the time
dependence is represented in terms of the ratio of $\overline\delta$,
the initial mean overdensity of linear perturbation at some $R$, to
the $\delta_{col}$, critical overdensity \cite{b16,b18,b17}, the
amplitude of perturbation which is to collapse at the moment
$t_{col}$.

\begin{figure}
  \begin{center}
    \includegraphics[width=8.0cm]{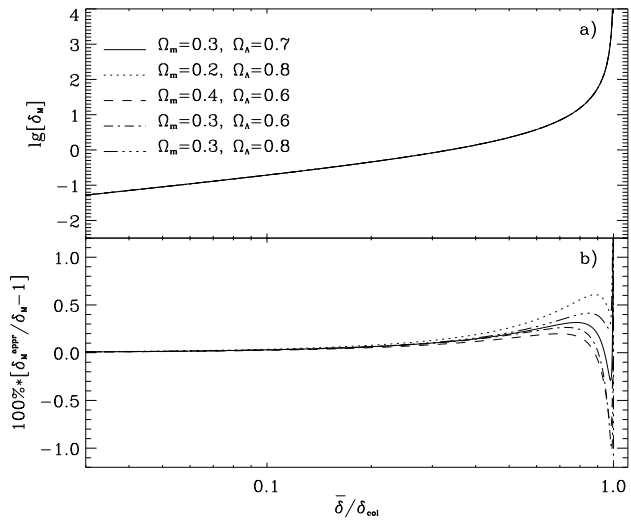}
  \end{center}
  \caption{Top panel: the dependence of non-linear amplitude of spherical
    perturbation on the linear one. Bottom panel: the accuracy of the
    approximation (\ref{approximation}) for some cosmological models, where 
    $\delta^{appr}_M$ denotes the right hand side of (\ref{approximation}) and 
    $\delta_M$ is the exact value.}
  \label{prf2}
\end{figure}

In the Fig.~\ref{prf2} the dependence of non-linear amplitude on the
ratio $\overline\delta/\delta_{col}$ is plotted, as computed by
integration of Eqs.\,(\ref{xscale}) and (\ref{Per2}).  The fit for the
dependence is simple,
\begin{eqnarray}
  \lg\left[1+\delta_M\right] &\simeq& -\delta_{col}\cdot\lg\left[1-\overline{\delta}/\delta_{col}\right] + A\cdot\lg^2\left[1-\overline{\delta}/\delta_{col}\right]\nonumber\\ 
  &+& B\cdot\lg^3\left[1-\overline{\delta}/\delta_{col}\right],
  \label{approximation}
\end{eqnarray}
the coefficients are $A = 0.0903$ and $B = 0.0074$.  This fit is similar to
the proposed by \cite{r01}, wherein the values $A=0$ and $B=0$ were assumed,
see also \cite{r02}. The fitting errors do not exceed 1\%
until the moment of complete collapse, when
$\delta_M\rightarrow\infty$ (bottom panel of Fig.~\ref{prf2}).

Given $\delta_M$, the density amplitude, $\delta_{\rho}$, is to be
evaluated at any radius $R$ with 
\begin{equation}
  \delta_{\rho} = \frac{1+\delta_M}{1+(\overline{\delta}-\delta)\frac{\partial}{\partial\overline{\delta}}\ln(1+\delta_M)}-1 .\label{delta_rho_appr}
\end{equation}

\subsection{Virialization and final parameters of individual spherical
  halo}

\begin{figure}
  \centerline{\includegraphics[width=7.0cm]{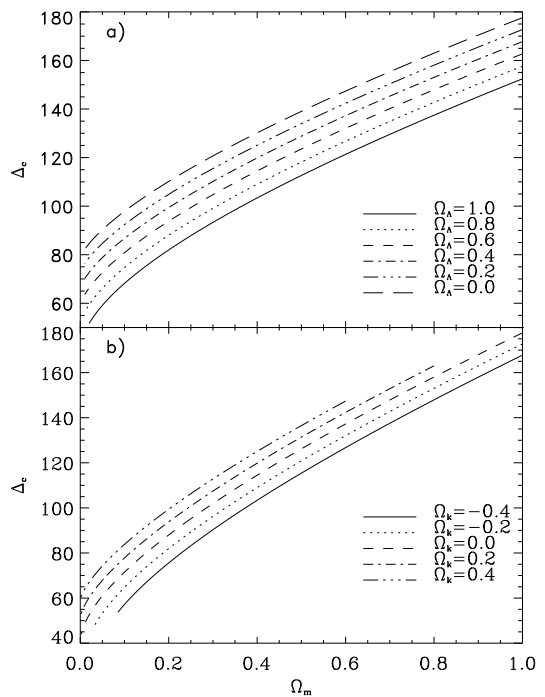}}
  \caption{The dependence of virialized spherical cloud density on
    $\Omega_m$ in units of the critical density at the moment of
    collapse, $\Delta_{vc}$, for models with fixed $\Omega_{\Lambda}$,
    (a), and $\Omega_K$, (b).}
  \label{Delta_c}
\end{figure}

Note that the true singularity at collapse stage in the center of the overdensity as a rule is not 
reached, since the falling of particles usually is not strictly radial and small-scale inhomogeneities 
within the cloud induce the  additional non-radial velocities of particles.
 The process of virialization is far from trivial, however, when the relaxation is finished 
 and dynamical equilibrium is established, the kinetic energy and the gravitational potential 
 satisfy the virial theorem. 
For instance, for a spherical relaxed halo
the kinetic energy per unit mass is determined by
$$
T_{vir}/m=\frac{1}{2}\left<v^2\right>_{vir} = \frac{1}{2} r\frac{\partial U_{vir}}{\partial r}.
$$

For $\Lambda$CDM model $U_{vir}=-H_0^2\Omega_{\Lambda}x_{vir}^2-H_0^2\Omega_m/x_{vir}$ \cite{b18}
and the total energy of isolated dark matter cloud is conserved. By
equating the total energy at turnaround point (kinetic energy is zero and $E_{tot}=U_{ta}$) to the one at virialization epoch
($E_{tot}=U_{vir}+T_{vir}$) we obtain
\begin{equation}
  2\Omega_{\Lambda}x_{vir}^2 +\frac{1}{2}\frac{\Omega_m}{x_{vir}}=\Omega_{\Lambda}x_{ta}^2+\frac{\Omega_m}{x_{ta}}.\label{xv-xta}
\end{equation}

With Eq.~(\ref{xscale}) for turnaround point, $\dot{x}(\tau_{ta})=0$,
we get the cubic equation for $x_{vir}$,
\begin{equation}
4\Omega_{\Lambda}x_{vir}^3+2\Omega_f x_{vir}+\Omega_m=0,
\label{x_vir} 
\end{equation}
with real root for overdensity ($\Omega_f<0$) in cosmology with
$\Omega_{\Lambda}>0$,  
\begin{equation}
  x_{vir} = \left(-\frac{2\Omega_{f}}{3\Omega_{\Lambda}}\right)^{\frac12} 
  \cos\left\{\frac13\arccos\left[-\frac{\Omega_m}{8\Omega_{\Lambda}}\left(-\frac{6\Omega_{\Lambda}}{\Omega_{f}}\right)^{\frac32}\right]-\frac{2\pi}{3}\right\}.
\label{xvir}
\end{equation} 

For $\Omega_{\Lambda}=0$, the Eq.~(\ref{xv-xta}) implies the strict
equality $x_{vir}=x_{ta}/2$. For $\Omega_{\Lambda}>0$,
$x_{vir}<x_{ta}/2$, albeit the difference is not large. In fiducial
model with $\Omega_{\Lambda}=0.7$ and $\Omega_m=0.3$ for perturbation
collapsing at current epoch, the relative difference
$(\frac12x_{ta}-x_{vir})/x_{vir}$ is indeed $\sim0.1$ and diminishes
with either $\Omega_{\Lambda}$ decrease or increase of collapse
redshift. Therefore, the approximation $x_{vir}\approx x_{ta}/2$ can
be applied in most cases.

Note, that the value of $x_{vir}$ depends on the local curvature
$\Omega_f$ and consequently on the collapse time, $t_{col}$. Also,
the virial mass density, $\rho_{vir} = \rho_m^0x_{vir}^{-3}$, depends
on $t_{col}$. It is convenient to represent the virial density in
units of critical one, taken at the moment of collapse:
\begin{equation}
\Delta_{vc}=\frac{\rho_{vir}(\tau_{col})}{\rho_{cr}(\tau_{col})}=
\frac{\Omega_mH_0^2}{x_{vir}^3(\tau_{col})H^2(\tau_{col})}.\label{Delta_c_exact}
\end{equation}
For the Einstein--de Sitter model ($\Omega_m=1$, $\Omega_{\Lambda}=0$)
this ratio does not depend on the collapse moment and equals
$\Delta_{vc}=18\pi^2\simeq178$. 

\section{Density profiles and concentration parameter}
\label{section3}


The basic assumption of halo model is that dark matter is
associated with virialized halos, which have some universal density
profile.  Halo density profile is described by generic expression,
$\rho(r) = \rho_s (r/r_s)^{-\gamma}(1+r/r_s)^{\gamma-\alpha}$, with
coefficients restricted by \cite{b73} to $2.5 \leq \alpha \leq 3$ and
$1 \leq \gamma \leq 1.5$.  The characteristic radius $r_s$ specifies
the distance, at which the slope of density profile changes. We use
universal NFW density profile \cite{b16} henceforth, this is a special
case of generic profile with values of slopes fixed as $\gamma=1$ and
$\alpha=3$:
\begin{equation}
  \rho(r) = \frac{\rho_s}{(r/r_s)(1+r/r_s)^2}.\label{rho_rm}
\end{equation}
For this density profile the total halo mass diverges logarithmically
with $r$, whence the size of each halo has to be limited to some
finite value.

The characterization of halo is a matter of convention. Here, the mass
of halo is defined as a mass of the whole matter contained within the volume
of radius $r_{vir}$. The quantity $r_{vir}$ is defined as a radius of
sphere, the mean internal density of which exceeds the value of the
critical density by some fixed factor. In case of the factor $200$ the halo mass
is denoted by $M_{200}$, for $\Delta_{vc}$ used as a factor the
$M_{\Delta}$ is a denotation of halo mass. Sometimes the factor is
assumed to be $180$, so that the $M_{180}$ is used accordingly. The index is
omitted when the choice of definition is clear from context.

The ratio of radius $r_{vir}$, used at defining the halo, to the
quantity $r_s$ is called the concentration parameter (or just
concentration) and denoted as $c$. Depending on the definition of
$r_{vir}$, the corresponding index is used as $c_{200}$,
$c_{\Delta}$, $c_{180}$. By defining the mass of halo and
concentration parameter, one defines the parameters of halo profile,
$\rho_s$ and $r_s$.

For halos of fixed mass the concentration is a stochastic variable,
with log-normal probability distribution function,
\begin{equation}
  p(c|m,z) dc= \frac{1}{\sqrt{2\pi} \sigma_{\ln c}}\exp\left[-\frac{\ln^2[c/\overline{c}(m,z)]}{2\sigma_{\ln c}^2}\right]d\ln c.
  \label{p_c}
\end{equation}
In such case the variance of concentration virtually does not depend
on the halo mass ($\sigma_{\ln c} = 0.2-0.35$, see \cite{b84}),
whereas the mean value of concentration depends on mass and
redshift. This dependence (the term ``mass dependence of
concentration'' is used hereafter) can be determined either by data of
simulations or derived analytically, see \cite{b16,b86,b5}. Since the
mass dependence follows from initial power spectrum of matter, the
analytical methods seem to be preferable. The changes in mass
dependence caused by modifications of shape or normalization of
initial power spectrum can be easily taken into account.

Analytical techniques aimed to study the mass dependences of profile
parameters are usually based on the treatment of \cite{b16}. The data
of simulations provide some indications of the growth of profile
specific density, $\rho_s$, with decrease of halo mass.  As it was
suggested in \cite{b16}, this is due to the tendency of less massive
but higher inhomogeneities to collapse earlier.  It was assumed also
that specific density of halo, $\rho_s$, is proportional to the matter
density of the Universe, taken for the moment of collapse,
\textit{i.e.}
\begin{equation}
  \rho_s  =  C\Omega_m\rho_{cr}(1+z_{col})^3,\label{rhos}
\end{equation}
with the proportionality constant $C$ to be determined from simulation.

The collapse time is defined in \textit{ad-hoc} manner. The collapse
is assumed to start at some moment of time, ``at which half the mass of the halo 
was first contained in progenitors more massive than some fraction
$f$ of the final mass'' \cite{b83}.

With Press-Schechter formalism this condition implies
\begin{equation}
  \mathrm{erfc}\,\left\{\frac{\delta_{col}(z_{col}|z)-\delta_{col}(z|z)}{\sqrt{2(\sigma^2(fM|z)-\sigma^2(M|z))}}\right\} = 
  \frac12, \label{erfc}
\end{equation}
leading to the equation:
\begin{eqnarray}
  \delta_{col}(z_{col}|z) &=& \delta_{col}(z|z) + C'\sqrt{\sigma^2(fM|z)-\sigma^2(M|z)}\nonumber\\ 
  &\simeq& \delta_{col}(z|z) + C'\sigma(fM|z),\label{zcol}
\end{eqnarray}
where $\sigma^2(M|z)=\sigma^2(M)(D(z)/D(0))^2$ and $\sigma^2(M)
\equiv \sigma^2(M|z=0)$.  Here, $z$ is the moment of halo observation,
$C'\approx 0.7$, the term $\sigma^2(M|z)$ was neglected in comparison
with $\sigma^2(fM|z)$. The values of $C$ and $f$ ought to be driven
from simulation data. In \cite{b16}, $f$ was found to be
virtually independent of cosmological parameters at $\approx 0.01$, meanwhile the coefficient $C
\sim 10^3$, and is strongly determined by the background
cosmological model and/or initial power spectrum.

As far as $C$ and $f$ have been determined, the specific density of
halo, $\rho_s$, can be estimated by Eqs.~(\ref{zcol}) and (\ref{rhos})
for given halo mass $M$ as well as other halo parameter, $r_s$.
Since there is no explicit analytical expression for dependence
$\delta_{col}(z_{col}|z)$ in such treatment, it ought to be recomputed
for each cosmology. Therefore, the following simplification of
Eq.\,(\ref{zcol}) had been proposed in \cite{b86}: 
\begin{equation}
  \sigma(fM_{\Delta}|z_{col}) = 1.686,\,\mbox{with}\,\,f = 0.01 . \label{zcol_bal}
\end{equation}
Also, the simple approximation was proposed therein for the
concentration, $c_{\Delta} = K(1+z_{col})/(1+z)$, with $K$
estimated from numerical modeling as $K=4$.

As it was shown in \cite{b5}, the above mentioned estimation of
concentration parameter from power spectra is applicable to CDM
cosmology. However, it is not a case for Warm Dark Matter (WDM),
since wrong dependences follow from it for smaller
scales. According to the computer simulations \cite{b5,r08}, for WDM the
concentration tends to grow with the mass increase, whereas for the CDM
the contrary dependence is expected.

Since the power spectrum and mass dependence of concentration share
the same behavior of slopes, it was proposed in \cite{b5} to replace
the Eqs.~(\ref{zcol}) and (\ref{zcol_bal}) by following:
\begin{eqnarray}
  &&\sigma_{\mbox{eff}}(M_s|z_{col}) = C_{\sigma}^{-1}, \nonumber \\ 
  &&\sigma_{\mbox{eff}}(M|z) = \sigma(M|z)\left(-\frac{d\ln \sigma(M)}{d\ln M}\right),
  \label{c_Eke}
\end{eqnarray}
where $C_{\sigma} \approx 28$ and the mass $M_s$ is the one
confined within the radius $r_{max} = 2.17r_s$, where the rotational
velocity of a particle has a maximum (for NFW profile). The following
estimation has been proposed in \cite{b5} for the concentration:
\begin{equation}
  c_{\Delta} = \left(\frac{\Delta_{vc}(z_{col})\cdot\Omega_m(z)}{\Delta_{vc}(z)\cdot\Omega_m(z_{col})}\right)^{1/3}\frac{1+z_{col}}{1+z}.
\label{c_Delta}
\end{equation}  
This approximation has an obvious drawback, as the moment of collapse, 
$z_{col}$, should be somehow known in advance, namely by numerical 
evaluation using iteration method for (\ref{c_Eke}).
Here, we propose to
eliminate these computations by altering a few basic assumptions. The
specific density of halo is assumed to be determined primarily by the
collapse of roughly homogeneous central region of protocloud. The
rest of the halo is formed afterwards around this core by the infall
of outer shells. The boundary of the core could be defined as the
point where the slope of density profile changes. Since the
accurate determination of such boundary is cumbersome, the mass of
core, $M_c$, can be estimated as the mass of halo contained within
radius $r_c=\beta r_s$, where $\beta$ will be estimated below. In other words, the value
of mean internal density of matter within the radius $r_c$ corresponds
to the density at the moment when dynamical equilibrium is
established.

According to these assumptions,
\begin{equation}
  M_c = \frac43\pi \rho_{vir}(z_{col}) r_c^3 =  \frac43\pi \Delta_{vc}(z_{col})\rho_{cr}(z_{col}) \beta^3r_s^3,
  \label{M_c1}
\end{equation}
where $\rho_{cr}(z_{col})$ denotes the critical density at the moment
of $z_{col}$. On the other hand, integration of the density profile
(\ref{rho_rm}) within $r_c$ yields
\begin{equation}
  M_c = 4\pi \rho_s r_s^3 \left[\ln(1+\beta)-\frac{\beta}{1+\beta}\right].
  \label{M_c2}
\end{equation}
Total mass of the halo is
\begin{eqnarray}
  M &=& 4\pi \rho_s r_s^3 \left[\ln(1+c)-\frac{c}{1+c}\right]\nonumber\\ &=& M_c \frac{\ln(1+c)-c/(1+c)}{\ln(1+\beta)-\beta/(1+\beta)},
  \label{M_Mv}
\end{eqnarray}
here $c$ is halo concentration.

In order to evaluate the specific density of halo $\rho_s$ the
condition of collapse (\ref{erfc}) should be redefined for the core of protohalo
of mass $M_c$ and radius $r_c$ at the observation moment.  
So, the new condition takes the form
\begin{equation}
  \frac{\delta_{col}(z_{col}|z)-\delta_{col}(z|z)}{\sqrt{2(\sigma^2(fM_c|z)-\sigma^2(M_c|z))}} = \mbox{const},
\end{equation}
where $f<1$. The term $\sigma^2(M)$
in Eq.~(\ref{zcol}) should not be neglected for accurate estimation of
concentration dependence on the slope and amplitude of power spectrum
(as in \cite{b5}). Moreover, it is crucial for the case of WDM because
$\sigma^2(M)$ changes slowly at small values of mass. So, we can rewrite the equation (\ref{zcol}) as power series in $(1-f)$ 
\begin{eqnarray}
  &&\delta_{col}(z_{col}|z)\approx   \delta_{col}(z|z) + C'\left[-\frac{d\sigma^2(M_c)}{d\ln M_c}(1-f)\right. \nonumber\\
  &&- \left.\frac12\frac{d^2\sigma^2(M_c)}{d\ln M_c^2}(1-f)^2  - \ldots\right]^{1/2}\frac{D(z)}{D(0)}. \label{zcolmy}
\end{eqnarray}
In the first order one can obtain
\begin{equation}
  \delta_{col}(z_{col}|z) \simeq \delta_{col}(z|z) + g\left[-\frac{d\sigma^2(M_c)}{d\ln M_c}\right]^{1/2}\frac{D(z)}{D(0)},
  \label{delta_zcol}
\end{equation}
where $g$ is a constant, the value of which can be drawn from simu\-lations.

To confront the Eq.~(\ref{delta_zcol}) with that of \cite{b5} the
approximation $\delta_{col}(z_{col}|z)=\delta_{col}(z_{col}|z_{col})
D(z)/D(z_{col}) \simeq 1.686D(z)/D(z_{col})$ is used, the term
$\delta_{col}(z|z)$ is neglected since for most of halos $z_{col}\gg z$ and thus
$\delta_{col}(z_{col}|z)\gg\delta_{col}(z|z)$.  With these assumptions, the Eq.~(\ref{delta_zcol}) is rendered
to
\begin{eqnarray}
  &&\frac{D(z_{col})}{D(0)}\sigma_{eff}(M_c) = (\sqrt{2} g/1.686)^{-1}, \nonumber\\ 
  &&\sigma_{\mbox{eff}}(M_c) = \sigma(M_c)\left(-\frac{d\ln \sigma}{d\ln M_c}\right)^{1/2}.\label{delta_zcol_transf}
\end{eqnarray}

The difference between Eqs. (\ref{delta_zcol_transf}) and
(\ref{c_Eke}) is apparent, namely the powers of derivatives, 1/2 in
(\ref{delta_zcol_transf}) versus 1 in (\ref{c_Eke}). Moreover, the
values of constants $\sqrt{2} g/1.686$ and $C_{\sigma}$ are not
necessary equal, as the masses $M_c$ and $M_s$ are defined by
different radii, $r_c= 0.5r_s$ and $r_{max} = 2.17r_s$
respectively. Hereafter, we advocate the use of
Eq.~(\ref{delta_zcol}) as more accurate and rigorously following from
condition of \cite{b16}.  

At the next step, to estimate the parameters of density profile of halo the
critical amplitude $\delta_{col}(z_{col}|z)$ should be linked with
relative density of virialized perturbation,
$\Delta_{vc}(z_{col})$.  For this $x_{vir}$ must be evaluated from Eq. (\ref{x_vir}) using the next
expression for local curvature
\begin{equation}
  \Omega_f(z_{col}) = \Omega_K - \frac53\Omega_m\frac{\delta_{col}(z_{col}|z)}{D(z)}\label{omega_fzcol}
\end{equation}
obtained from Eq.~(\ref{delta2}).
Further, the parameter of halo density profile $r_s$ as a function of $M_c$ can
be found by evaluating the critical amplitude
$\delta_{col}(z_{col}|z)$ for given mass $M_c$ with
(\ref{delta_zcol}) along with Eqs.~(\ref{x_vir}) and
(\ref{omega_fzcol}), above-mentioned definitions and Eq.~(\ref{M_c1}):
\begin{equation}
  r_s = \frac{x_{vir}}{\beta}\left(\frac{(M_c/10^{12}h^{-1}M_{\odot})}{1.163\Omega_m}\right)^{1/3},
\end{equation}
where $r_s$ has dimension Mpc/h and we taken into account that $4\pi\rho_{cr}(0)/3 \simeq 1.163 \cdot 10^{12} M_{\odot}h^{-1}/(\mathrm{Mpc}/h)^3$.
The specific density of halo, $\rho_s$, is evaluated from
Eq.~(\ref{M_c2}).  

As long as the ratio of mean density of halo to specific density is a
function of concentration  $c$,
\begin{equation}
  \frac{\rho_{halo}}{\rho_s} = \frac{3M}{4\pi r_{vir}^3\rho_s} = \frac{3}{c^3}\left[\ln\left(1+c\right)-\frac{c}{1+c}\right],
\end{equation}
the concentration is a function of that ratio, the approximation expression for which is given in \cite{b7}:
\begin{equation}
\label{eq:66}
  c  \simeq \left[\frac23\frac{\rho_{halo}}{\rho_s} + \left(\frac{1.1}{2.0}\frac{\rho_{halo}}{\rho_s}\right)^{0.387}\right]^{-1}.
\end{equation}

It is convenient to express the specific and mean densities of halo in
units of critical density, \textit{i.e.} $\Delta_{sc} \equiv
\rho_s/\rho_{cr}$ and $\Delta_{hc} \equiv \rho_{halo}/\rho_{cr}$, or
in units of mean density of matter, as $\Delta_{sm} \equiv
\rho_s/\overline{\rho}_m$ and $\Delta_{hm} \equiv
\rho_{halo}/\overline{\rho}_m$ correspondingly. Thus,
${\rho_{halo}}/{\rho_s} = {\Delta_{hc}}/{\Delta_{sc}} =
{\Delta_{hm}}/{\Delta_{sm}}$. According to the condition, used to
define the halo radius $r_{vir}$, one of the values, either
$\Delta_{hc}$ or $\Delta_{hm}$, should be constant for all halos,
meanwhile either $\Delta_{sc}$ or $\Delta_{sm}$ is evaluated by
formulas:
\begin{eqnarray}
  \Delta_{sc} & = & \frac{\Omega_m}{x_{vir}^3}\frac{H_0^2}{H^2(z)}\frac{\beta^3/3}{\ln(1+\beta)-\beta/(1+\beta)}, \label{eq:67} \\
  \Delta_{sm} & = & \frac{1}{x_{vir}^3}\frac{1}{(1+z)^3}\frac{\beta^3/3}{\ln(1+\beta)-\beta/(1+\beta)}.
\label{Delta_sc}
\end{eqnarray}
Then the total halo mass can be simply evaluated using Eq.~(\ref{M_Mv}).
\begin{figure}
  \begin{center}
    \includegraphics[width=8.0cm]{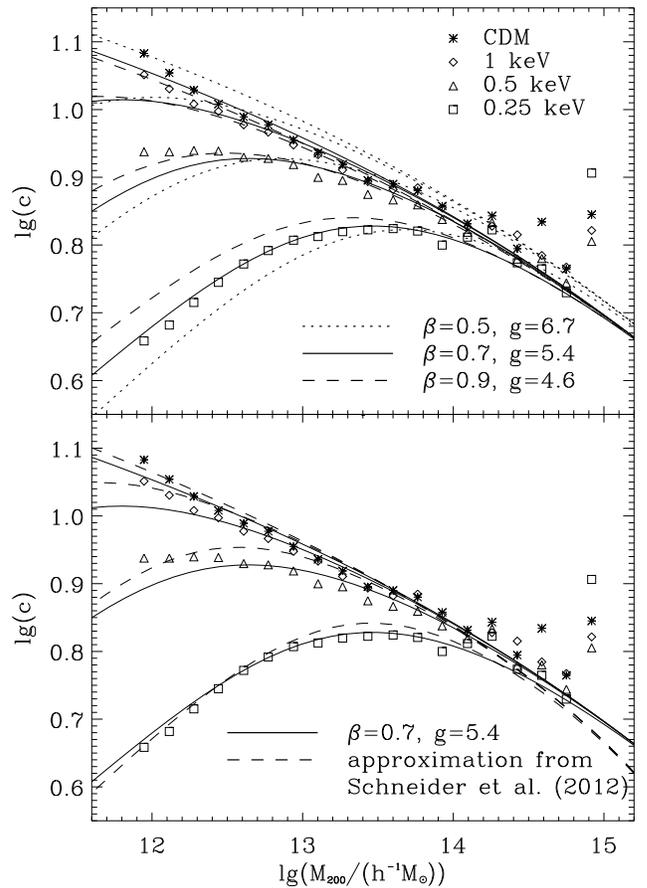}
  \end{center}
  \caption{The dependences of concentration parameter $c$ on halo mass $M_{200}$ for different cosmologies. 
  The data of simulations for CDM (stars) and
    WDM with different masses of dark matter particles ($m = 1$~keV - diamonds, $m = 0.5$~keV - triangles,  $m
    = 0.25$~keV - squares) are taken from \cite{r08}. Top panel: our approximation for different
    parameters, $\beta=0.5$, $g=6.7$ -- dotted lines; $\beta=0.7$,
    $g=5.4$ -- solid lines; and $\beta=0.9$, $g=4.6$ -- dashed
    lines. Bottom panel: the comparison of our approximations
    ($\beta=0.7$, $g=5.4$, solid lines) with approximations in \cite{r08} (dashed lines).}
\label{warm_c}
\end{figure}

The approximations for dependences of concentration on mass are
presented in Fig.~\ref{warm_c} for CDM and WDM (for set of DM
particle masses) along with the data of simulations carried out by
\cite{r08}. Also, we used the data of the simulations to find the
best-fit values for the parameters, $\beta=0.7$ and $g=5.4$, and
plotted them along with the approximation of same authors for
comparison, see bottom panel.
All calculations were performed for a number of
$\Lambda$CDM and $\Lambda$WDM cosmologies with parameters
$\Omega_m=0.2726$, $\Omega_{\Lambda}=0.7274$, $h=0.704$,
$\sigma_8=0.809$ and $n_s=0.963$.

The values of halo concentrations correlate with halo ages, 
so that the oldest halos are expected to have larger concentrations (see \cite{r07} 
for details).
According to hierarchical CDM scenario of clustering,
the halos of lower masses should be formed in first turn, therefore
they should be of larger concentrations. Meanwhile, for the WDM the
perturbations at small scales are suppressed by free-streaming. As a
result, in case of WDM the low-mass halos are mainly formed after
cooling of warm dark matter caused by expansion of the Universe, hence
they appear to have smaller values of concentration.

\begin{figure}
  \begin{center}
    \includegraphics[width=8.0cm]{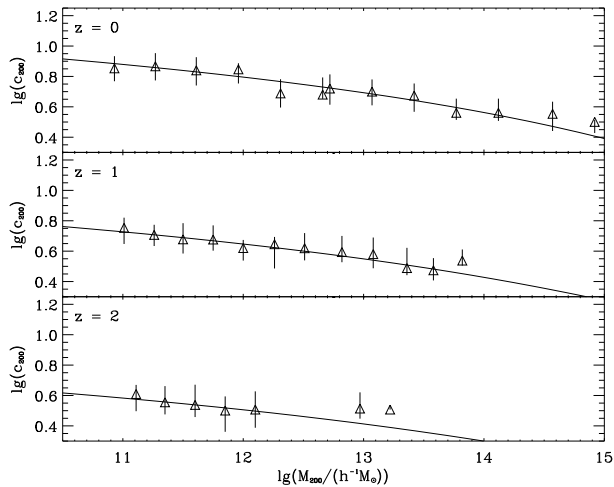}
  \end{center}
  \caption{The dependence of concentration, $c_{200}$, on the mass
    $M_{200}$. The triangles represent the modeling by \cite{b3}. The
    solid lines represent our results. The plots are given for the
    redshifts $z = 0,1,2$ in downward order.}
\label{c_wmap5}
\end{figure}

Another comparison of our predictions with simulations is presented
in Fig.~\ref{c_wmap5}, this time with respect to redshift evolution.
The evaluated dependence of halo concentration, $c_{200}$, on mass
$M_{200}$, is presented therein along with the modeling data from
\cite{b3}, the parameters of cosmological model are taken from the 5-year
Data Release of WMAP \cite{b88}: $\Omega_{\Lambda} = 0.721$, $\Omega_m
= 0.279$, $\Omega_b = 0.0441$, $h=0.719$, $\sigma_8=0.796$ and
$n_s=0.963$.  Three plots represent dependences for the set of
redshifts, $z = 0,1,2$. Quite good agreement is seen between our
calculations and the data of simulations at all redshifts.

\section{Mass functions of halos and matter power spectrum}

The pioneering paper of Press and Schechter \cite{b34} introduced
an analytical approach to statistical description for galaxy clusters
distribution. The model of spherical collapse underpins this
formalism, the halos are associated with the peaks of an initial
Gaussian field of density perturbations. This Press-Schechter
formalism utilizes the halo mass function to describe the distribution
of halos over masses.  The approach was refined and extended
afterwards in \cite{b10,b36,b37} to allow for the merger histories of
dark matter halos. The process of halo merging is assumed to be
hierarchical at the large scales and described with characteristic
collapsing mass scale, $m(t_{col})$, complemented with r.m.s. of
density perturbations, $\sigma(m)=\delta_{col}(t_{col})$.
This mass grows with time through merging of halos and should asymptotically 
approach in distant future the limit $m_{\infty}$ at which 
$\sigma(m_{\infty}) = \delta_{min}$, where $\delta_{min}$ is the minimal amplitude of 
linear density perturbations which can reach the turnaround point followed by collapse 
and formation of virialized objects for cosmologically justified time 
(see for details \cite{b17}).
For cosmology with $\delta_{min}=0$ the clustering of dark matter never ends in sense
that all halos of the Universe will merge in far future.

\subsection{Halo mass function}
\label{MassFunction}

According to \cite{b10}, the Press-Schechter mass function $n(m,z)$,
\textit{i.e.}  the number density of gravitationally bound objects
with masses $m$ at redshift $z$, is supposed to satisfy the condition
\begin{equation}
  \nu F(\nu) \equiv \frac{m^2n(m,z)}{\overline{\rho}_m}\frac{d\ln m}{d\ln \nu}=\sqrt{\frac{\nu}{2\pi}}\exp\left\{-\nu/2\right\},
  \label{PressShehter}
\end{equation}
where $\nu \equiv (\delta_{col}(t_{col})/\sigma(m))^2$ and
$\overline{\rho}_m$ is the background matter density.
 
The Press-Schechter mass function is proven to be qualitatively
correct, however in some details the discrepancies with the
data of N-body simulations are found. Therefore, the number of improvements to this approach are
proposed. For instance, the treatment of the
collapsing perturbations as ellipsoidal rather than spherical
diminishes the discrepancies (see \cite{b38}). Indeed, by assuming the
average ellipticity of perturbation with mass $m$ and amplitude
$\delta$ to be ${e}_{mp} = (\sigma(m)/\delta)/\sqrt{5}$, a simple
relation was obtained in \cite{b8} to connect ellipsoidal and
spherical collapse thresholds
\begin{equation}
  \delta_{ec}(m,t_{col}) = \delta_{col}(t_{col})\left(1+0.47\left[\frac{\sigma(m)}{\delta_{col}(t_{col})}\right]^{1.23}\right).
  \label{delta_ec}
\end{equation}

Also, the excursion set model was used in \cite{b10} to estimate the
mass function associated with ellipsoidal collapse,
\begin{equation}
  \nu F(\nu) = A(p)\left(1+\nu^{-p}\right)\sqrt{\frac{\nu}{2\pi}}\exp\left\{-\nu/2\right\},
  \label{ChetTormen}
\end{equation}
where parameter $p \simeq 0.3$ and function $A(p) \equiv
\left[1+2^{-p}\Gamma(1/2-p)/\sqrt{\pi}\right]^{-1} \simeq 0.3222$ are
determined by requirement that the whole mass is gathered within halos,
\textit{i.e.} the integration of $F(\nu)$ over $\nu$ yields unity. In
order to match the data of GIF numerical simulations the mass function
(\ref{ChetTormen}) has been parameterized in \cite{b38} as
\begin{equation}
  \nu F(\nu) = A(p)\left(1+(q\nu)^{-p}\right)\sqrt{\frac{q\nu}{2\pi}}\exp\left\{-q\nu/2\right\}.
  \label{ChetTormen_new}
\end{equation}
The additional parameter $q$ was found to be $q=0.707$, later it was
re-determined in \cite{b44} to be $q=0.75$. The ellipsoidal threshold
for such mass function, estimated in the framework of the excursion set approach \cite{b8}, is as follows:
\begin{equation}
  \delta_{eq}(m,t_{col}) = {q}^{\frac12}\delta_{col}(t_{col})\left(1+0.5\left[\frac{\sigma(m)}{q^{\frac12}\delta_{col}(t_{col})}\right]^{1.2}\right).
  \label{delta_eq}
  \end{equation}

  Two different algorithms are commonly used to identify the dark matter
halos within data of numerical N-body simulations: the
friend-of-friend (FOF) algorithm \cite{b39} and the spherical
overdensity (SO) finder \cite{b40}. The FOF procedure depends on just
one free parameter, $b$, which defines the linking length as
$b\overline{n}^{-1/3}$, where $\overline{n}$ is the average density of
particles. Thus, in the limit of very large number of particles per
halo, FOF approximately selects the halo as matter enclosed by an
isodensity surface at which $\rho = \overline{\rho}/b^3$.  SO
algorithm finds the values ​​of the average halo density in spherical
volumes of various sizes. The criterion for halo identification is
equality of the average density over the sphere to certain value
$\kappa\overline{\rho}_m$, where $\overline{\rho}_m$ is mean density
of matter in sample and $\kappa$ is a parameter of algorithm. For the
NFW density profile these algorithms are not to be identical as they
lead to different mass dependences of the halo concentration parameter
$c$.  Nevertheless, the similarity of halo mass functions was found in
\cite{b41} using SO ($\kappa = 180$) and FOF ($b=0.2$) halo finders.

Hereafter, we refer to halo as a gravitationally bound system which
has reached the state of dynamical equilibrium, meanwhile both SO and
FOF finders select the groups of close particles regardless of their
dynamical properties.  To divide such halos into virialized (relaxed)
and non-virialized parts, it was suggested in \cite{b42} to assess the
dynamical state of each halo processed by FOF algorithm by means of
three objective criteria: 1) the substructure mass function $f_{sub}$,
2) the center of mass displacement $s=|r_c-r_{cm}|/r_{vir}$ and 3)
the virial ratio $2T/U$. In \cite{b2} the r.m.s. of the NFW fit
to the density profile have been used too.

As far as virial density $\rho_{vir} = \rho_{cr}\Delta_{vc} =
\overline{\rho}_m\Delta_{vm}$, it seems appropriate to use SO
halo-finder with $\kappa = \Delta_{vm} = \Delta_{vc}/\Omega_m$. The
equality $\kappa \simeq 180$ is valid for any redshift in flat
$\Omega_m=1$ cosmology (this is close to the $b\simeq 0.2$ for FOF
algorithm), meanwhile for $\Lambda$CDM cosmology with $\Omega_m=0.3$,
$\Omega_{\Lambda}=0.7$ the quantities $\kappa$ and $b$ depend on
redshift: $\kappa \simeq 97/0.3 \simeq 324$ ($b\simeq 0.164$) at $z=0$
and slowly decrease (increase) to the limit $\kappa \simeq 180$
($b\simeq 0.2$) at high $z$.  However, as it is shown in \cite{b41},
the shape of mass function is invariant if we simply identify clusters
with a constant linking length, $b = 0.2$, for all redshifts and
cosmologies.

\begin{figure}
  \begin{center}
    \includegraphics[width=9.0cm]{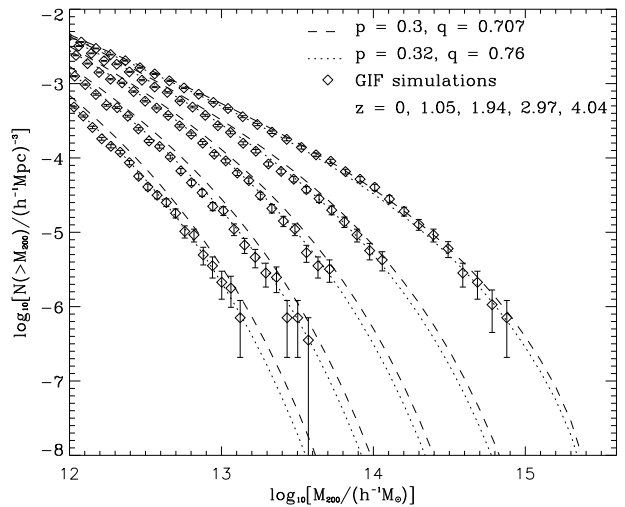}
  \end{center}
  \caption{The halo mass function for different redshifts ($z = 0,
    1.05, 1.94, 2.97, 4.04$ from top to bottom). The dashed lines are Sheth-Tormen
    approximation with parameters $p=0.3$ and $q=0.707$ \cite{b38},
    the dotted lines show the same approximation with modified
    parameters $p=0.32$ and $q=0.76$, diamonds show results from N-body numerical simulations performed by
    GIF/Virgo collaboration \cite{b46}.}
  \label{mf_gif}
\end{figure}

The halo mass function derived from N-body simulations of GIF/Virgo
collaboration is plotted in Fig.~\ref{mf_gif}.  The catalogs of halos
were built from simulations and made
available\footnote{http://www.mpa-garching.mpg.de/GIF}. For each halo
detected by FOF-algorithm ($b=0.2$) the catalogs include the mass
$M_{200}$, confined within the central part of halo with overdensity
$\Delta_{vm} = 200$ (see \cite{b46} for details). The mass rescaling
slightly affects the 'observed' mass function. We have re-determined
the parameters of Sheth-Tormen approximation to be $p=0.32$ and
$q=0.76$. As it follows from Fig.~\ref{mf_gif}, the refined parameters
provide a better fit for data than ones from \cite{b38}, namely
$p=0.3$ and $q=0.707$. The mass function is defined here as a number
density of halos with masses exceeding the specified mass $m$,
\begin{equation}
  N(>m) = \int\limits_m^{\infty}n(m',z)dm' = \int\limits_{m}^{\infty}\frac{\overline{\rho}_m}{m'}\nu F(\nu)\frac{d\ln\nu}{dm'}dm'.
  \label{NM}
\end{equation}
Note, that variations in FOF or SO halo finder parameters also alter
the total number of detected halos, meanwhile the mass rescaling
influences the shape of mass function, not the total number of halos.

The Press-Schechter formalism \cite{b34} implies that halos are
shaped out of regions with initial overdensities
$\delta\geq\delta_{col}$, \textit{i.e.} the collapsed ones. However,
this does not prevent the initially lower overdensities, $\delta <
\delta_{col}$, to reach the value $\Delta_{vm}$. For the non-linear
overdensity $\delta_M = 180$ the corresponding
initial amplitude of density perturbation (in the units of critical one) is $q^{\frac12}
\equiv \overline{\delta}/\delta_{col} \simeq 0.95$, as it follows from (\ref{approximation}). Thus, the
SO-algorithm ($\kappa=\Delta_{vm}=\delta_M+1$) can by chance mark as halos the
non-virialized regions, the initial amplitude of which exceeds $q^{\frac12}\delta_{col}$ but is less than 
$\delta_{col}$ (for spherical overdensities).

It seems reasonable to assume that the elliptical 'q-threshold'
$\delta_{eq}$ is directly connected to spherical 'q-threshold'
$q^{\frac12}\delta_{col}$ through (\ref{delta_eq}) just in the same
manner as the elliptical collapse threshold $\delta_{ec}$ is related
to the spherical collapse threshold $\delta_{col}$ with
(\ref{delta_ec}). However, the estimate obtained above, $q \simeq
0.95^2 \simeq 0.90$ for $\Delta_{vm} = 180$, substantially deviates
from $q \simeq 0.75$, estimated by numerical simulations using FOF
($b=0.2$) and SO ($\kappa=180$) algorithms.  The large halos are
supposed to be close to spherical, so their mass distribution should
comply to Press-Schechter one.  More interesting, when $\delta_{col}$
in (\ref{PressShehter}) is replaced by the spherical 'q-threshold'
$0.95\delta_{col}$ ($0.9\nu$ for $\nu$), a good match to numerical
simulations is attained for large $\nu$ and therefore large masses.
However, the Press-Schechter mass function tends to overestimate the
number of halos with smaller masses, because low-mass proto-halos are
more elliptical and therefore according to (\ref{delta_ec}) need
larger initial amplitude to became a halo.

\subsection{The dark matter power spectrum}

A luminous object is determined by clumping of baryon matter, which in
turn is tightly governed by gravitational potential of dark matter.
Whence, the observable spatial distribution of galaxies should follow the 
distribution of dark matter, since the latter dominates by density.
So, in order to reconstruct the observable distribution of galaxies
the characteristics of distribution of dark matter are needed.

\subsubsection{Two-point correlation function and power spectrum of
  discrete and continuous distributions}

In statistics, the inhomogeneity of spatial distribution is usually
described either by the two-point correlation function or by its
Fourier transform, the power spectrum. The latter can be directly
drawn by Fourier transformation of relative density fluctuations. In the
case of continuous distribution it is
\begin{equation}
  \delta(\vec r) = (2\pi)^{\frac{3}{2}}V^{\frac{1}{2}}\int\delta_{\vec k}e^{-i\vec k\vec r}d^3\vec k
  = \frac{(2\pi)^{\frac{3}{2}}}{V^{\frac{1}{2}}}\sum\limits_{\vec k} \delta_{\vec k}e^{-i\vec k\vec r},\label{delta_rk}
\end{equation}
here $V$ denotes `volume of periodicity' to be properly chosen.  The
coefficients of (\ref{delta_rk}) are:
\begin{equation}
  \delta_{\vec k} = \frac{1}{(2\pi)^{3/2}V^{1/2}}\int\delta(\vec r)e^{i\vec k\vec r}d^3\vec r
  = \frac{V^{\frac{1}{2}}}{(2\pi)^{\frac{3}{2}}}\sum\limits_{\vec r} \delta(\vec r)e^{i\vec k\vec r}.\label{delta_kr}
\end{equation}

The Fourier amplitude, squared and averaged over the different
directions of vector $\vec k$, yields the power spectrum,
$\mathcal{P}(k) = \left<|\delta_{\vec k}|^2\right>$.  The two-point
correlation function is readily derived from given power spectrum,
\begin{eqnarray}
  && \xi(r) = \left<\delta(\vec r')\delta(\vec r'+\vec r)\right> = \frac{(2\pi)^3}{V}\sum\limits_{\vec k} \left<|\delta_{\vec k}|^2\right> e^{i\vec k\vec r} \nonumber\\
  && =\int d^3\vec k \left<|\delta_{\vec k}|^2\right>e^{i\vec k\vec r} = 4\pi\int\limits_{0}^{\infty}k^2dk\mathcal{P}(k)\frac{\sin(kr)}{kr},
\end{eqnarray}
as well as variance of the amplitude within the sphere of radius $R$,
\begin{eqnarray}
  \sigma^2(R) &=& \left<\delta_R^2\right> = 4\pi\int k^2\mathcal{P}(k)W^2(kR)dk \nonumber\\
  &=& \int \Delta^2(k)W^2(kR)d\ln k,
\end{eqnarray}
where $W(x)=3(\sin(x)-x\cos(x))/x^3$ is a window function for sphere,
the quantity $\Delta^2(k) = 4\pi k^3\mathcal{P}(k)$ is a
``dimensionless'' power spectrum.

The power spectrum is evaluated from correlation function as
\begin{equation}
  \mathcal{P}(k) = \frac{1}{(2\pi)^{3}}\int d^3\vec r e^{-i\vec k\vec r}\xi(r).\label{Pk_neper}
\end{equation}

The galaxy catalogs (and the data of numerical simulations) involve
discrete distributions of objects (``particles''). Thus, the equations
(\ref{delta_rk}) and (\ref{delta_kr}) should be rewritten with
$\rho(\vec r) =\sum\limits_im_i\delta_D(\vec r-\vec r_i)$, where $m_i$
is the mass of $i$-th particle, $\delta_D(\vec r-\vec r_i)$ is
three-dimensional Dirac function,
\begin{eqnarray}
  \delta_{\vec k}=\frac{1}{(2\pi)^{3/2}\left<m\right>\overline{n}V^{1/2}}\sum\limits_im_ie^{-i\vec k\vec r_i},
\end{eqnarray}
$\overline{n}$ is spatially averaged number density of particles,
$\left<m\right> = \sum m_i/(\overline{n}V)$ is the mean mass.

The relation of power spectrum to correlation function is provided in
\cite{Peebles1993}, there
\begin{equation}
  \mathcal{P}(k) = \frac{\left<m^2\right>}{(2\pi)^{3}\overline{n}\left<m\right>^2} + \frac{1}{(2\pi)^{3}}\int d^3\vec r e^{-i\vec k\vec r}\xi(r),\label{Pk_diskr}
\end{equation}
with $\left<m^2\right> = \sum m_i^2/(\overline{n}V)$. The first term
in right-hand side is a shot noise, denoted henceforth by
$\mathcal{P}_{shot}$. It is inherent for discrete distribution and
caused by finiteness of the number density of particles
$\overline{n}$. At $\overline{n}\to \infty$, \textit{i.e.} for
continuous distribution, the Eqs.~(\ref{Pk_diskr}) and (\ref{Pk_neper})
converge. The second term in right-hand side of (\ref{Pk_diskr}) is
denoted henceforth as $\mathcal{P}_{\xi}(k)$ to emphasize the
non-random (correlated) nature of distribution. Thus, the
Eq.~(\ref{Pk_diskr}) can be written in more compact form as
$\mathcal{P}(k) = \mathcal{P}_{shot} + \mathcal{P}_{\xi}(k)$.

\subsubsection{Non-linear power spectrum in halo model}

Within halo model the distribution of matter is treated in a mixed,
discrete-continuous manner. The distribution of spatially separated
halos of different mass is considered and the distribution of matter
within each halo is described by continuous density profile
(\ref{rho_rm}). Therefore, the power spectrum is split into two terms,
one to describe the distribution of halos and the second to describe
the distribution of matter within individual halo. The splitting can
be derived rigorously taking into account that Fourier amplitudes of
density perturbations are (see for details Appendix \ref{app_a})
\begin{equation}
  \delta_{\vec k} = \frac{1}{\overline{\rho}}\int\limits_0^{\infty} m\cdot n(m)\delta_{\vec k}(m)\overline{y}(m,k)dm .
\end{equation}

The $\overline{\rho} = \overline{\rho}_m^0/a^3$ is an average matter
density at the moment of time determined by scale factor
$a=(1+z)^{-1}$ and $n(m)$ is the number density of halos with mass $m$
in comoving coordinates, estimated with (\ref{ChetTormen_new}).

The function $\overline{y}(m,k)$ is a Fourier transform of density
profile (\ref{rho_rm}) expressed explicitly by analytical form
\begin{eqnarray}
  &&\overline{y}(m,k)  = \frac{4\pi}{m} \int\limits_{0}^{r_{vir}/a}\frac{\sin(kR)}{kR}\rho(Ra) R^2 dR\nonumber\\
  && = \frac{4\pi\rho_sr_s^3}{ma^3}\left\{ \left[Si\left(\frac{kr_s}{a}(1+c)\right)-
 Si\left(\frac{kr_s}{a}\right)\right]\sin\left(\frac{kr_s}{a}\right)\right.\nonumber\\ 
  &&+\left. \left[Ci\left(\frac{kr_s}{a}(1+c)\right)-Ci\left(\frac{kr_s}{a}\right)\right]\cos\left(\frac{kr_s}{a}\right) 
  \right.\nonumber\\ 
  &&- \left.\frac{a}{(1+c)kr_s}\sin\left(c\frac{kr_s}{a}\right)\right\},\label{y_dm}
\end{eqnarray}
where $c$ is the halo concentration, $\rho_s$ and $r_s$ are parameters
of density profile, $Si(x)$ and $Ci(x)$ are integral sine and cosine
respectively.  The halo profile depends on the physical coordinates,
while the power spectrum is associated with comoving coordinates as
$R=r/a$.  The term $4\pi\rho_sr_s^3/m$ can be expressed via halo
concentration parameter $c$ using the Eq.~(\ref{M_Mv}).

The power spectrum $\mathcal{P}(k|m,m')$ of spatial distribution of
halos with given masses $m$ and $m'$ is following:
\begin{eqnarray}
\mathcal{P}(k|m,m')&=&\frac12\left<\delta_{\vec k}^{*}(m)\delta_{\vec k}(m')+\delta_{\vec k}(m)\delta_{\vec k}^{*}(m')\right>\nonumber\\
&=&\frac{\delta_{m,m'}}{(2\pi)^{3}n(m)} + \mathcal{P}_{\xi}(k|m,m'). \nonumber
\end{eqnarray}
where $\delta_{m,m'}$ is Kronecker symbol, $\mathcal{P}_{\xi}(k|m,m')$
is Fourier image of two-point cross-correlation function of the halos,
the angle brackets in the right-hand side denote the averaging over
the directions of $\vec k$.

After the series of mathematical transformations we obtain
\begin{eqnarray}
  \mathcal{P}(k) &=& \frac{1}{(2\pi)^3\overline{\rho}^2}\int\limits_0^{\infty} 
  m^2\cdot n(m)|\overline{y}(m,k)|^2dm \nonumber\\
  &+& \frac{1}{\overline{\rho}^2}\int\limits_0^{\infty} m\cdot n(m)\overline{y}(m,k)dm\label{power_halo1}\\
  &&\times\int\limits_0^{\infty} m'\cdot n(m')\overline{y}(m',k)dm'\mathcal{P}_{\xi}(k|m,m').\nonumber
\end{eqnarray}
The quantity $n(m)$ is a number density of halos of masses $m$.

Under the assumption of linearity the cross-correlation power spectrum can be
represented as $\mathcal{P}_{\xi}(k|m,m') \approx
b(m)b(m')\mathcal{P}_{lin}(k)$, $\mathcal{P}_{lin}(k)$ is the
linear power spectrum of spatial distribution of matter, $b(m)$ is the
biasing parameter which characterizes the skew between distributions of
halos and matter.

The requirement of homogeneity at largest scales imposes that
expression (\ref{power_halo1}) has to asymptotically approach zero for
small wave numbers $k$. Nevertheless, the first term in right-hand
side of (\ref{power_halo1}) never diminishes, because the binning of
matter into separate halos (a kind of discretization) introduces the
noise into the procedure.  The expression for noise is derived from
the first term in (\ref{power_halo1}) by letting the distribution of
halo matter to be homogeneous and substituting of Fourier image of
profile $\overline{y}(k,m)$ by window function $W(kR)$, where
$R=(3m/(4\pi\overline{\rho}_m))^{1/3}$. After noise elimination the
final expression for the power spectrum of spatial distribution of
dark matter is following:
\begin{eqnarray}
  && \!\!\!\!\! \!\!\!\!\! \!\!\!
  \mathcal{P}(k) = \frac{1}{(2\pi)^3}\int\limits_0^{\infty} 
  \left(\frac{m}{\overline{\rho}}\right)^2 n(m)\left[|\overline{y}(m,k)|^2-W^2(kR)\right]dm \nonumber\\
  &&\, + \left[\int\limits_0^{\infty} \frac{m}{\overline{\rho}} b_1(m) n(m)\overline{y}(m,k)dm\right]^2\mathcal{P}_{lin}(k).
\label{power_halo2}
\end{eqnarray}

In accordance with \cite{r10} the factor
$[|\overline{y}(m,k)|^2-W^2(kR)]$ is used instead of
$[\overline{y}(m,k)-W(kR)]^2$, as mentioned in the review
\cite{b0} on the halo model. It should be stressed, that at
quasilinear stage it yields rather small deviations from the numerical
simulation\footnote{At quasilinear stage the shape of power spectrum
  is still mainly determined by the shape of initial power spectrum,
  yet already differs from it (see \cite{b14} for details). The halo
  model tends to underestimate the power at quasilinear stage
  (\cite{b0}), in comparison with numerical simulations.} because
$[|\overline{y}(m,k)|^2-W^2(kR)] \ge [\overline{y}(m,k)-W(kR)]^2$ at
$k\sim 1/R$.

\begin{table}
  \caption{The parameters of Large Box (LB) and GIF2 simulations, available from Max Planck Institute for Astrophysics in Garching (http://www.mpa-garching.mpg.de).}
  \begin{center}
    \begin{tabular}{c|c|c|c|c}
      Simul. &  Npar & L (Mpc/h) & $m_p$ ($M_{sun}$/h) & $l_{soft}$ (Kpc/h)\\
      \hline
      GIF2 &  400$^3$ & 110.0 & 1.73$\times10^9$ & 6.6\\
      LB &  512$^3$ & 479.0 & 6.86$\times10^{10}$ & 30\\
    \end{tabular} 
  \end{center}
  \label{tab}
\end{table} 

With Eq.~(\ref{power_halo2}) the power spectrum of dark matter is
computed for the broad range of scales to confront our estimations
with results of Large Box and GIF2 N-body simulations available from Max
Planck Institute for Astrophysics in Garching.  The results of
simulation are released as files with coordinates, velocities and
identification numbers of particles.  The parameters of simulations,
namely the total number of particles, the size of box, the mass of
particles (assumed equal for all particles) and the scale of smoothing
are presented in the Tab.~\ref{tab}. The latter is introduced in order
to eliminate numerical singularities due to particles proximity, when
floating-point errors are difficult to control.

To reproduce the structure at small scales a simulation should engage
the high number density of particles.  On the other hand, large volume
is required to reproduce properly the structure at large scales. A
pursuit to simultaneously meet both requirements leads to the huge
numbers of particles and consequently to the enormous amount of
computational efforts. So, the commonly used trick is to run separate
simulations for largest and smallest scales.  Whence, the Large Box
simulations cover large volumes, whereas the the GIF2 simulations
provide us with data for small scales with larger number density of
particles.

\begin{figure}
  \begin{center}
    \includegraphics[width=8.5cm]{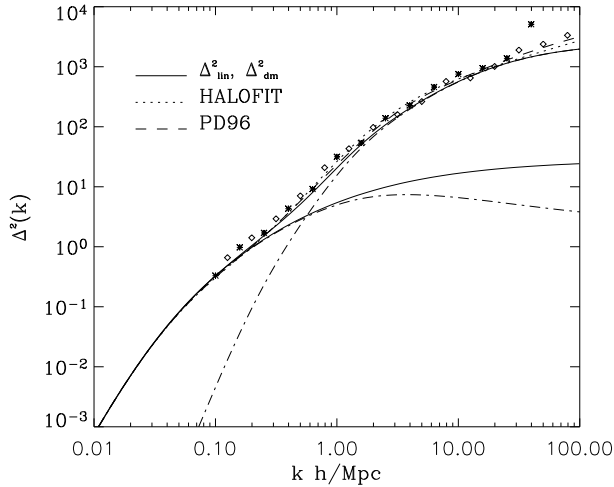}
  \end{center}
  \caption{Dark matter power spectrum from LB (asterisk) and GIF2
    (diamonds) simulations. Solid lines represent the primordial
    linear power spectrum from \cite{b49} (lower line) and our
    predictions for non-linear one (upper line). Dash-dotted lines
    show the 'halo-halo' and the 'shot noise' components. Dotted and
    dashed lines represent PD96 \cite{b15} and HALOFIT \cite{b14}
    approximations respectively. The parameters of $\Lambda$CDM model here are as follows:
    ($\Omega_m$, $\Omega_{\Lambda}$, $h$, $\sigma_8$, $n_s$)=(0.3,
    0.7, 0.7, 0.9, 1).}
  \label{power_Dk}
\end{figure}

The power spectrum was evaluated by computation of the
sums, $Sr(\vec k) = \sum_i \cos(\vec k \vec{r}_i)$ and $Si(\vec k) = \sum_i
\sin(\vec k \vec{r}_i)$, followed by overall summation,
$\mathcal{P}(\vec k) = (Sr^2(\vec k) +
Si^2(\vec k) )/((2\pi)^{3}\overline{n}^2V)$. The final power spectrum was estimated by
averaging over directions of $\vec k$. To eliminate the noise, the power spectrum of homogeneous
distribution was computed in advance and substracted later from the
total power spectrum.

In Fig.~\ref{power_Dk} the results are presented for different
techniques. The dark matter power spectrum predicted by our halo
model, Eq.~(\ref{power_halo2}), apparently matches the LB/GIF2
non-linear power spectrum through all scales up to $k\sim100$
h/Mpc.  Also, the PD96 \cite{b15} and HALOFIT \cite{b14}
approximations are plotted therein, based on the halo model of
Hamilton et al. \cite{Hamilton1991}, as well as scaling relations and
fits to numerical simulations.  All these approximations appear to
properly fit the LB/GIF2 non-linear power spectrum at the whole range
of scales. The linear power spectrum was evaluated by analytical
approximation from \cite{b49} (lower solid line in
Fig.~\ref{power_Dk}) and normalized to $\sigma_8=0.9$.

Since the non-linear corrections are not essential at $k\le0.2$ h/Mpc,  
the power spectrum appears to be linear there (Fig.~\ref{power_Dk}).
The non-linear clustering 
enhances the power spectrum at smaller scales, $k>0.2$ h/Mpc. 
Both approximations, our 
(\ref{power_halo2}) and HALOFIT, reproduce such behavior
appropriately. Consistency of our estimation with numerical
simulations data and HALOFIT approximation proves the plausibility of
our approach.

The halo mass function in WDM cosmology is expected to decline at low
masses as $n_h(m) = (1+m_{hm}/m)^{-0.6}n_{ST}(m)$ \cite{r08}, where
$n_{ST}(m)$ denotes the Sheth-Tormen mass function described in
section \ref{MassFunction}. The WDM tends to clump less, so that 
contributes largely to a smooth component of density field, $\rho_s$,
with $\overline{\rho} = \overline{\rho}_h + \overline{\rho}_s$
\cite{r11,r08}.  
To treat the WDM within the framework of halo model  
the separate particles of dark matter are considered as point
halos with mass $m_{DM}$, immersed into smooth component.
Thus, the total number density of halos is
\begin{equation}
 n(m) = n_{h}(m) + \frac{\overline{\rho}_s}{m_{DM}}\delta_D(m-m_{DM}).
\end{equation}
With substitution to (\ref{power_halo2}), similarly to \cite{r11}, the
power spectrum
\begin{eqnarray}
  && \!\!\!\!\! \!\!\!\!\! \!\!\!
  \mathcal{P}(k) = \frac{1}{(2\pi)^3}\int\limits_0^{\infty} 
  \left(\frac{m}{\overline{\rho}}\right)^2 n_h(m)\left[|\overline{y}(m,k)|^2-W^2(kR)\right]dm\nonumber\\
  &&\,  + \left[\int\limits_0^{\infty} \frac{m}{\overline{\rho}} b_1(m) n_h(m)\overline{y}(m,k)dm+b_s\frac{\overline{\rho}_s}{\overline{\rho}}\right]^2\mathcal{P}_{lin}(k).\nonumber
\label{power_halo2_wdm}
\end{eqnarray}
where biasing factor of smooth component can be obtained from 
\begin{equation}
 b_s\frac{\overline{\rho}_s}{\overline{\rho}} = 1-\int\limits_0^{\infty} \frac{m}{\overline{\rho}} b_1(m) n_h(m)dm.\nonumber
\end{equation}

\begin{figure}
  \begin{center}
    \includegraphics[width=8.5cm]{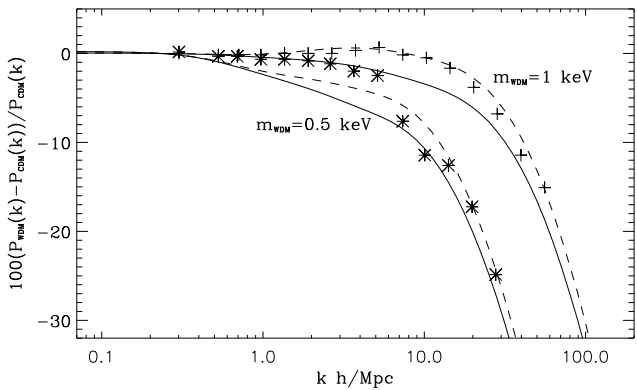}
  \end{center}
  \caption{The difference (per cent) between non-linear power spectra
    of $\Lambda$CDM and $\Lambda$WDM models with ($\Omega_{m}$,
    $\Omega_{\Lambda}$, $\Omega_b$, $h$, $n_s$, $\sigma_8$)=(0.2711,
    0.7289, 0.0451, 0.703, 0.966, 0.809) and two values of WDM
    particle mass: $m_{wdm} = 1$~keV and $m_{wdm} = 0.5$~keV.}
  \label{power_warm}
\end{figure}

The applicability of these formulas was verified by comparison with
the results of numerical simulations from \cite{b100}.  The initial
distribution of warm dark matter particles for simulation was
generated by following linear power spectrum: 
\begin{equation}
  \mathcal{P}_{lin}^{(wdm)}(k) = \mathcal{P}_{lin}^{(cdm)}(k)\left[\left(1+(\alpha k)^{2\nu}\right)^{-5/\nu}\right]^2,
 \label{PWDM1}
\end{equation} 
with $\nu=1.12$. The parameter $\alpha$ (in units of Mpc/h) depends on the mass of~WDM
particles $m_{wdm}$, their density $\Omega_{wdm}$ and Hubble
parameter as
$$\alpha(m_{wdm}) = 0.049\left(\frac{1keV}{m_{wdm}}\right)^{1.11}
\left(\frac{\Omega_{wdm}}{0.25}\right)^{0.11}\left(\frac{h}{0.7}\right)^{1.22}$$
(see also \cite{Hansen2002} and \cite{Viel2005}).

The non-linear power spectrum of matter density perturbations at $z$ =
0.5 was evaluated by (\ref{power_halo2}) for $\Lambda$CDM and
$\Lambda$WDM cosmologies with parameters set ($\Omega_{m}$,
$\Omega_{\Lambda}$, $\Omega_b$, $h$, $n_s$, $\sigma_8$)=(0.2711,
0.7289, 0.0451, 0.703, 0.966, 0.809) and two masses of warm dark
matter particles, $m_{wdm}$, 1 and 0.5~keV, the same as in
\cite{b100}.  The Fig.~\ref{power_warm} represents the relative
discrepancies (percentage) between non-linear power spectra of cold
and warm dark matter (solid lines), also the corresponding spectra
from simulations are plotted along (Fig.~7 in \cite{b100}). Our
results reveal qualitative consistency with simulations however
quantitative differences are still noticeable.

It is worth to mention the discrepancy of halo model and numerical
simulations in case of warm dark matter \cite{Smith2011}, see the
bottom panels of Figure 7 in paper \cite{b100} (green line). That
estimation appears to be suppressed in comparison with numerical
simulation and seems to be closer to our results. 
The plausible explanation is that the low-mass halos are more 
clustered in WDM models than in CDM ones.
As it was noted in \cite{r12}, ``formation of low mass
halos almost solely withing caustic pancakes or ribbons connecting
larger halos in a 'cosmic web' '', and ``voids in this web are almost
empty of small halos, in contrast to the situation in CDM theory''.
This leads to larger values of biasing at $m<m_{hm}$ in WDM models
with respect to CDM \cite{r08}.  We assume that the reason why small
halos with mass below $m_{hm}$ are so strongly clustered is that they
belong (at least partially) to some larger halos (i.e. they are
satellites). Dashed lines in Fig.~\ref{power_warm} represent the
computations for the case when masses of all halos are increased by
4\%, to be above $m_{hm}$.

Note some aspects of the problem to be addressed in further studies:
\begin{itemize}
\item The halos with mass $< m_{hm}$ can appear in result of: i) tidal
  stripping of dark matter from initially more massive halos, ii)
  evaporation of subhalos from large mass halos and iii) clustering
  in cold component of dark matter\footnote{The dark matter particles
    are collisionless, whence part of them, having small velocities,
    can be considered as a cold dark matter.}. Clarification of the
  contribution of each of such mechanisms is needed to update
  properly the halo model.
\item When the power spectra were calculated, the variance of the
  parameter of halo concentration $\sigma_{\ln c}=0.25$ was assumed to
  be the same for the cold and warm dark matter and independent of
  the halo mass. It follows from Fig.~\ref{power_warm}, that the
  deviations can be caused by the halo concentration variations.
\item Halo model by itself has a number of problems and not to be
  considered as ultimately accurate. It is based on some strong
  assumptions, contains a series of approximations and uncertain
  statistical procedures, thus prone to systematic errors.
\end{itemize}

\subsection{The galaxy power spectrum}

As the baryon gas falls into potential wells of virialized dark matter
halos and subhalos it is heated up to virial temperature of $T =
\frac12 \mu m_pv_{vir}^2/k_B \approx 2\times10^4 \mu_{0.6}
(M/10^8M_{\odot})^{2/3}[(1+z)/10]$~K, where $\mu = 0.6\mu_{0.6}$ is the
mean molecular mass of post-shock gas and $M$ is the mass of halo or
subhalo progenitors. The temperature of baryon matter gradually
decreases afterwards due to the cooling processes (see \cite{b4}). It
results in fragmentation to smaller clumps with Bonnor-Ebert mass,
$M_{BE} \simeq 700M_{\odot}(T/200
\textrm{K})^{3/2}(n_b/10^4\textrm{cm}^{-3})^{-1/2}$, where $n_b$ is
the total number density of baryon particles.  At the final stage of
this fragmentation the stars and galaxies are formed (see \cite{b4}
and \cite{b6} for details). Since formation of galaxies is driven by
gravity of dark matter, the spatial distribution of galaxies should
track the spatial distribution of dark matter. In other words, the
fluctuations of dark matter density $\delta_{DM}(\vec r) =
\rho_{DM}(\vec r)/\overline{\rho}_{DM} - 1$ correlate with
fluctuations of galaxy number density $\delta_g(\vec r) = n_g(\vec
r)/\overline{n}_g - 1$.  

In halo model, the galaxy number density fluctuations have the following
Fourier amplitude:
\begin{equation}
  \delta_{g|\vec k} \; = \frac{1}{\overline{n}_g}\int\limits_{m_{min}}^{\infty} \left<N_g|m\right> n(m)\delta_{\vec k}(m)\overline{y}_g(k,m)dm
  \label{deltak_gal}
\end{equation}
(see Appendix \ref{app_b} for details ). Here $\left<N_g|m\right>$ is
a mean number of galaxies in halo with mass $m$, $\overline{y}_g(k,m)$
is a Fourier transform of galaxy number density profile and $m_{min}$
denotes the lowest limit for halo mass below which no galaxies are
formed. Such limit naturally stems from degrading efficiency of star
formation in halos of low mass \cite{b1} and conditions imposed on the
sample of galaxies (see \cite{b26,b27}).

The considerations of previous subsection are summarized in the following
galaxy power spectrum:
\begin{eqnarray}
  && \!\!\!\!\! \!\!\!\!\! \!\!\!
  \mathcal{P}_{g}(k) = \frac{1}{(2\pi)^3}\int\limits_{m_{min}}^{\infty} 
  \left(\frac{\left<N_g|m\right>}{\overline{n}_g}\right)^2 n(m)\nonumber\\
  &&\!\!\!\!\! \!\!\!\!\! 
  \times\left[|\overline{y}_{g}(m,k)|^2-W^2(kR_g)\right]dm \label{galaxypk}\\
  &&\!\!\!\!\! \!\!\!\!\! 
  + \left[\int\limits_{m_{min}}^{\infty} \frac{\left<N_g|m\right>}{\overline{n}_g} b_1(m) n(m)\overline{y}_g(m,k)dm\right]^2\mathcal{P}_{lin}(k),
  \nonumber
\end{eqnarray}
where $R_g = (3\left<N_g|m\right>/(4\pi\overline{n}_g))^{1/3}$.  The
resulting equation is similar to the corresponding expression for the
galaxy power spectrum from \cite{b0}. The difference is caused by
elimination of the noise as described above. Also, the term ${\left
    <N_g|m\right>}^2$ has been obtained instead of
$\left<N_g(N_g-1)|m\right>$ in \cite{b0}. For large-mass halos the
term $\left<N_g|m\right>$ is large, and it seems appropriate to assume
the probability distribution $p(N_g|m)$ to be one of Poisson. In this
case ${\left<N_g|m\right>}^2 \approx
\left<N_g(N_g-1)|m\right>$. However, such approximation is not valid
for low-mass halos.

As it has been outlined in \cite{b20}, galaxies within halo usually
are disposed around the center (central galaxy) and within each of its
subhalos (satellite galaxies). This gives a clue how to find out the
distribution of galaxies over the halo and how it is connected to the
substructure.  Massive halos usually undergo the violent relaxation,
so the resulting velocity dispersion does not depend on masses of
particles or subhalos. Therefore for the number density of the
satellites within halo the following equation is appropriate:
\begin{eqnarray}
  && \!\!\!\!\! \!\!\!\!\! \!\!\!
  n_{sg}(r) = \sum_{m\geq m'_{min}}\!\!\!\!\!  n_{sh}(m,r) = \sum_{m\geq m'_{min}}\!\!\!\!\! n_0(m)\exp \left\{-\frac{m\Phi(r)}{kT}\right\}\nonumber\\
  && \;\, = n_{g}^0 \exp \left\{-\frac{3\Phi(r)}{\sigma_v^2}\right\} = \frac{n_{g}^0}{\rho_s}\rho(r),
  \label{galhalo_1}
\end{eqnarray}
where $n_{sh}(m,r)$ denotes the dependence of the number density of
halo particles (subhalos) of mass $m$ on the radial distance.

As above, we assume that galaxies are formed within subhalos with
masses $m \geq m'_{min}$, where $m'_{min}$ is less than $m_{min}$
because subhalos usually lose the mass due to tidal deprivation of
their outskirts.  Baryon matter (stars) is concentrated to the center
and more tightly bound, meanwhile dark matter is stripped off. Thus, a
subhalo at the time of observation is apparently a poor tracer of
potential well, which still determines galaxy properties such as
stellar mass or luminosity.  A better tracer is the subhalo mass at
the time when it falls into the host halo or its maximal mass over
its history \cite{r20,b20}. For massive halos with numerous satellites the presence
of the central galaxy can be neglected.  In such case, as follows from
(\ref{galhalo_1}), the assumption $\overline{y}_g(m,k) \simeq
\overline{y}(m,k)$ is correct.  This result agrees with \cite{b102},
there the spatial distribution of satellites is studied using SDSS
spectroscopic and photometric galaxy catalogs. They found that
satellite profiles generally have a universal form well-fitted by
NFW approximation.

However, as long as low-mass halos possess a small number of galaxies,
slow relaxation can be important as well. In result, the profile of
satellite galaxy number density generally deviates from the profile of
dark matter density. However, such discrepancy is difficult to detect
because of large statistical uncertainties in determination of the
profile of the galaxy number density in such halos. Let us note that
in this case the presence of central galaxy could not be discarded.

\begin{figure*}
  \begin{minipage}{175mm}
    \begin{center}
      \leavevmode
      \includegraphics[width=14.0cm]{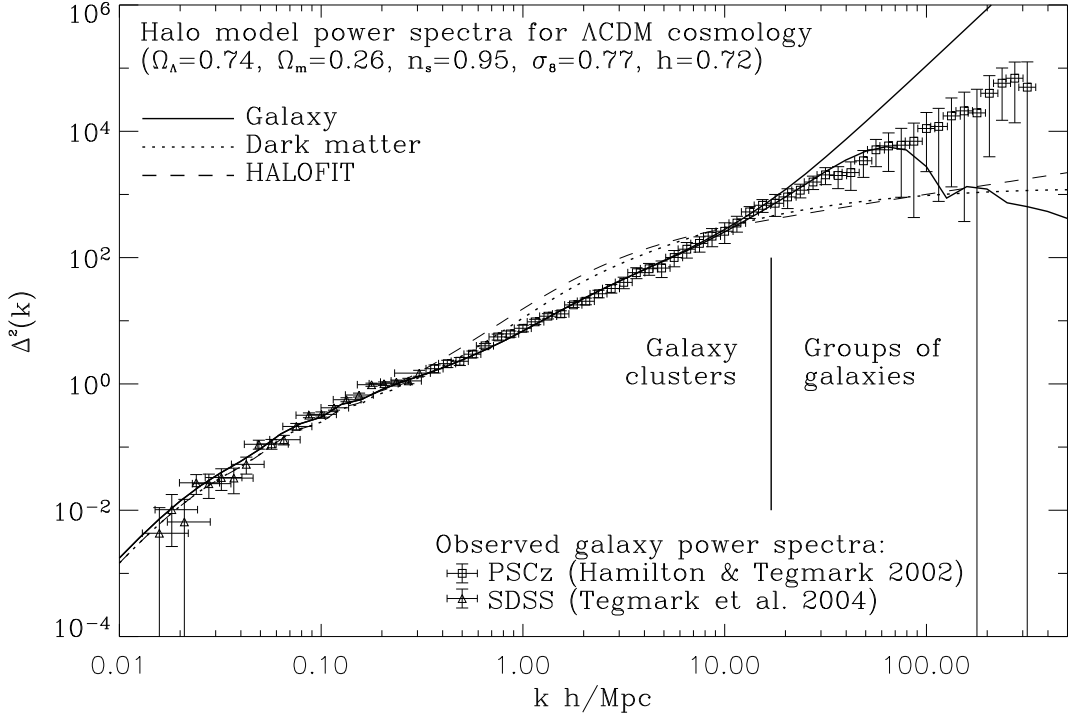}
    \end{center}
    \caption{The power spectra of galaxies (solid line) and dark matter
      (dotted line) calculated in halo model for $\Lambda$CDM
      cosmology with ($\Omega_m$, $\Omega_{\Lambda}$, $h$, $\sigma_8$,
      $n_s$) = (0.26, 0.74, 0.72, 0.77, 0.95). The squares and
      triangles represent the observed galaxy power spectrum from PSCz
      \cite{b26} and SDSS \cite{b27} galaxy catalogs respectively.}
    \label{GalaxyDk}
  \end{minipage}
\end{figure*}

The spatial number density of the galaxies is a sum of the halo and
subhalo number densities: $n_g(\vec r) = n_h(\vec r) + n_{sh}(\vec
r)$. The spatial fluctuation of galaxy number density can be thereby
split into fluctuations of halo and subhalo numbers densities:
\begin{eqnarray}
  \delta_g(\vec r) &=& \frac{n_h(\vec r) + n_{sh}(\vec r)}{\overline{n}_h+\overline{n}_{sh}} - 1\nonumber\\ 
  &=&  \frac{1}{\overline{n}_h+\overline{n}_{sh}}\left[\overline{n}_h\delta_h(\vec r)+\overline{n}_{sh}\delta_{sh}
  (\vec r)\right],
\end{eqnarray}
where as before the overlines denote averaging in space. Corresponding
Fourier amplitude takes the form
\begin{eqnarray}
  && \!\!\!\!\! \!\!\!\!\! \!\!\!
  \delta_{g|\vec k} =  \frac{1}{\overline{n}_h+\overline{n}_{sh}}\left[\overline{n}_h\delta_{h|\vec k}+\overline{n}_{sh}
  \delta_{sh|\vec k}\right]\label{power_gal_new1}\\
  && \!\!\!\!\! \!\!\!\!\! \!\!\!
  = \frac{1}{\overline{n}_h+\overline{n}_{sh}}
  \int\limits_{m_{min}}^{\infty}\!\!\!\!\!\left[1 + \left<N_{sh}|m\right>\overline{y}_{sh}(k,m)\right] n(m)\delta_{\vec k}(m)dm,\nonumber
\end{eqnarray}
where $\left<N_{sh}|m\right>$ is the average number of subhalos
confined within the halo of mass $m$, virtually the number of
satellites. Since the average number of galaxies accounts for central
galaxy and satellites, $\left<N_{g}|m\right> = 1 +
\left<N_{sh}|m\right>$.  By comparing Eqs.~(\ref{deltak_gal}) and
(\ref{power_gal_new1}) one can obtains
\begin{equation}
  \overline{n}_{g}=\overline{n}_h+\overline{n}_{sh} = \int\limits_{m_{min}}^{\infty}\!\!\!\!\!\left[1 + 
  \left<N_{sh}|m\right>\right] n(m)dm,\nonumber
\end{equation}
and
\begin{equation}
  \overline{y}_{g}(k,m) = \frac{\left<N_{sh}|m\right>\overline{y}_{sh}(k,m)+
  \overline{y}_{c}(k,m)}{\left<N_{sh}|m\right> + 1},\nonumber
\end{equation}
where $\overline{y}_{sh}(k,m)$ is a Fourier image of the subhalo
number density profile, meanwhile $\overline{y}_{c}(k,m)$ is a Fourier
image of the probability of finding the central galaxy within the
halo.

In cases of strictly central location of `core' galaxies in all halos
with masses $m$ $\overline{y}_{c}(k,m)=1$.  For massive halo,
$\left<N_{sh}|m\right> \gg 1$, so $\overline{y}_{g}(k,m) \simeq
\overline{y}_{sh}(k,m)$, whereas for low-mass halo,
$\left<N_{sh}|m\right> \ll 1$, $\overline{y}_{g}(k,m) \simeq
\overline{y}_{c}(k,m)$. As follows from Eq.~(\ref{galhalo_1}) for
massive halos we can assume $\overline{y}_{sh}(k,m) \simeq
\overline{y}(k,m)$.  For simplicity, let us extend this approximation to
the case of low-mass halo. It should not bring significant errors to
$\overline{y}_{g}(k,m)$ because in the case of $\left<N_{sh}|m\right>
\ll 1$ the core galaxy is dominating, so $\overline{y}_{g}(k,m) \simeq
\overline{y}_{c}(k,m)$.

To specify the dependence $\left<N_{sh}|m\right>$ and to provide a
direct link to the galaxy sample the CLF \cite{b19,b32,b33} or CMF
\cite{b20}) can be used.  The CLF, $\Phi(L|m)dL$, yields the average
number of galaxies with luminosity $L\pm dL/2$ which reside within a
halo of mass $m$. The CMF, $\Phi(m_*|m)dm_*$, yields the average
number of galaxies with stellar masses in the range $m_*\pm dm_*/2$
which reside within halo of mass $m$. The CMF (as well as CLF) can be
split into central (core) and satellite parts so that $\Phi(m_*|m)
= \Phi_s(m_*|m) + \Phi_c(m_*|m)$. This allows us to calculate the
average number of satellites with a stellar masses exceeding $m_*$
within the halo with mass $m$ (see \cite{b20} for details),
$$
\left<N_{sh}|m,m_*\right> = \int\limits^{\infty}_{m_*}\Phi_s(m_*'|m)dm_*',
$$
and the probability of finding the appropriate central galaxy is
$$
\left<N_{c}|m,m_*\right> = \int\limits^{\infty}_{m_*}\Phi_c(m_*'|m)dm_*',
$$
where upper limit is assigned to infinity, although it actually does
not exceed the halo mass $m$.  To calculate the power spectrum of
galaxies, we assume that $\left<N_{g}|m\right> =
\left<N_{sh}|m,m_*\right> + \left<N_{c}|m,m_*\right>$ and
\begin{equation}
  \overline{y}_{g}(k,m) = \frac{\left<N_{sh}|m,m^*\right>\overline{y}_{sh}(k,m)+\left<N_{c}|m,m^*\right>\overline{y}_{c}(k,m)}{\left<N_{sh}|m,m^*\right> + \left<N_{c}|m,m^*\right>}.
  \label{y_sh_c}
\end{equation}
The average number of galaxies with a stellar mass larger than $m_*$
is given by
\begin{equation}
  \overline{n}_{g} = \int\limits_{0}^{\infty}\left[\left<N_{sh}|m,m_*\right> + \left<N_{c}|m,m_*\right>\right] n(m)dm.
\end{equation}
The similar calculations are valid for CLF. Hence, the halo model
describes the connection between galaxy power spectrum and stellar
masses or luminosities of the sample of galaxies.

Note that our approach differs from the one proposed in \cite{b0} since it allows to consider the 
displacements of position of central galaxy in halos.
This is important for small mass halos which tend to have large ellipticity and shallow 
potential wells. So, we predict that halos which contain single galaxy give contribution to the 1st 
term of galaxy power spectrum  (\ref{galaxypk}), also called as 1-halo term. 

To prove our approach, we calculate the galaxy power spectrum along with error bars 
using the CMF from \cite{b20}
for $\Lambda$CDM cosmology with parameters ($\Omega_m$,
$\Omega_{\Lambda}$, $h$, $\sigma_8$, $n_s$) = ($0.26$, $0.74$, $0.72$,
$0.77$, $0.95$). The initial dark matter power spectrum,
$\mathcal{P}_{lin}(k)$, was computed with the CAMB code
\cite{camb,camb_source} for $\Omega_b=0.05$.  The galaxy power
spectrum was evaluated by the Eq.~(\ref{y_sh_c}) and
Eq.~(\ref{galaxypk}) with $W(kR_g)$ replaced by
\begin{eqnarray}
  &&\lim_{\mathcal{P}(k)\rightarrow 0}\overline{y}_{g}(k,m) = \nonumber\\
  &&\frac{\left<N_{sh}|m,m^*\right>W(kR_{s})+\left<N_{c}|m,m^*\right>W(kR_c)}{\left<N_{sh}|m,m^*\right> + \left<N_{c}|m,m^*\right>},
  \nonumber
\end{eqnarray}
where $R_{s} =
\left(3\left<N_{sh}|m,m^*\right>/(4\pi\overline{n}_g)\right)^{1/3}$
and $R_c = (3\left<N_{c}|m,m^*\right>/(4\pi\overline{n}_g))^{1/3}$.
Also, it is assumed
\begin{equation}
 \overline{y}_{sh}(k,m) = \int\limits_{all c} \overline{y}(k,r_s,c')p(c'|m,z)dc'.
\end{equation}
where $\overline{y}(k,r_s,c)$ denotes the dependence (\ref{y_dm}) and
$p(c|m,z)$ is the probability distribution function for concentration
(\ref{p_c}) with variance $\sigma_{\ln c} = 0.25$.

The obtained galaxy and dark matter power spectra are presented in
Fig.~\ref{GalaxyDk} along with observed galaxy power spectra from
PSCz \cite{b26} and SDSS \cite{b27} galaxy catalogs.

The upper solid line represents the assumption that the core galaxies
in all halos with masses $m$ are located strictly in their centers, so
$\overline{y}_{c}(k,m)=1$.  The lower solid line represents the result
for assumption that central galaxies are homogeneously distributed
over the spherical volume of radius $1.1 r_s$, so
$\overline{y}_{c}(k,m)=W(1.1r_sk)$.  We define the lower limit on the
stellar masses of the galaxies to be $m_{*} = 5\times
10^6$~$M_{\odot}$.

Thus, at large scales, $k\le 1$ h/Mpc, the dark matter and
galaxy power spectra coincide, at galaxy cluster scales, $1\le k\le
20$ h/Mpc, they are close and start to diverge at smaller
scales, $k>20$ h/Mpc, where luminous matter is substantially
more clustered than dark matter.

\section{Conclusions}

The presented semi-analytical treatment is our implementation of halo
model and it is proven to be correct in describing and interpretation
of the clustering of the matter at the non-linear stage of evolution,
both in simulated and observed Universe. Some of basic elements of
theory are reviewed and improved to calculate the dark matter and
galaxy power spectra.

A new technique is proposed for calculating halo concentration
parameter, $c$, with phenomenology of halo merging, density profiles
and statistical properties taken into account. The simple expression
for estimation (\ref{eq:66}) depends on the relation of the halo
overdensity, $\Delta_{hc}$ or $\Delta_{hm}$, and corresponding
characteristic halo overdensity, $\Delta_{sc}$ or $\Delta_{sm}$
respectively.  This relation is evaluated without computing redshift
of halo collapse, $z_{col}$, by set of equations: (\ref{eq:66}),
(\ref{delta_zcol}), (\ref{omega_fzcol}), (\ref{xvir}) and
(\ref{eq:67}) or (\ref{Delta_sc}) as well.  Such technique has been
applied to calculate the concentration parameter for $\Lambda$CDM and
$\Lambda$WDM cosmological models and the concordance with data of
simulations \cite{b5,b3} for vast range of halo masses
(Figs. \ref{warm_c}, \ref{c_wmap5}) has been revealed.

The parameters of Sheth-Tormen approximation for halo mass function
were re-evaluated as $p=0.32$ and $q=0.76$ (see Fig. \ref{mf_gif}) to
provide best-fit to the data of GIF/Virgo N-body simulations
\cite{b46} (see Fig. \ref{mf_gif}). 

This modified and extended halo model enables to predict the dark
matter and galaxy power spectra at small scales up to $k\sim 100$
h/Mpc by means of semi-analytical methods:
Eqs.~(\ref{power_halo1}), (\ref{power_halo2}), (\ref{galaxypk}). The
estimated spectra agree with non-linear power spectra determined
from Large Box and GIF2 N-body simulations (Fig.~\ref{power_Dk}) as
well as with estimations by galaxy catalogs PSCz \cite{b26} and SDSS
\cite{b27} (Fig. \ref{GalaxyDk}). Moreover, with the assumption on
presence of the central galaxies in all halos with masses $m$
($\overline{y}_{c}(k,m)=1$) the technique predicts galaxy power
spectrum matching well the observational one up to $k\sim 20$
h/Mpc. Meanwhile, when non-central position of most massive
galaxies in halos is assumed, ($\overline{y}_{c}(k,m)=W(1.1r_sk)$),
the predictions agree with the observational data up to $k\sim 80$
h/Mpc.

The calculated non-linear galaxy power spectrum for $\Lambda$CDM
cosmology with ($\Omega_m$, $\Omega_{\Lambda}$, $h$, $\sigma_8$,
$n_s$) = (0.26, 0.74, 0.72, 0.77, 0.95) corresponds to the
observational one for lower limitation on the stellar masses of the
galaxies $m_{*} = 5\cdot 10^6$~$M_{\odot}$.  To attain the same level
of agreement of the predicted galaxy power spectrum with extracted
from galaxy surveys at smaller scales ($k>80$ h/Mpc), a new,
much more complicated approach for the formation of groups of galaxies
should be elaborated.

Despite the ambiguities in the definition of halo, determining of
their mass, concentration and substructure, halo model provides a good
reproduction of such characteristics of large-scale structure of the
Universe as the power spectrum and correlation function of the spatial
distribution of dark matter and galaxies. In this paper we have shown how
relation between statistics of the dark matter clustering
obtained from numerical simulations and galaxy statistics obtained
from large galaxy surveys allows to calculate the power spectrum of
the spatial distribution of the galaxies.

\section*{Acknowledgments}
 
This work was supported by the project of Ministry of Education and Science of Ukraine (state registration number 
0113U003059), research program ``Scientific cosmic research'' of the National Academy of Sciences of Ukraine 
(state registration number 0113U002301) and the SCOPES project No. IZ73Z0128040 of Swiss National Science Foundation. 
Authors also thank to A. Boyarsky and anonymous referees for useful comments and suggestions.

\appendix

\section[A]{Fourier modes of dark matter density inhomogeneities}
\label{app_a}

The power spectrum can be derived in more rigorous manner by the
series of following mathematical transformations:
\begin{widetext}
\begin{eqnarray*}
  &&\delta_{\vec k} = \frac{1}{(2\pi)^{\frac32}V^{\frac12}}\int \limits_{V} \delta(\vec r) e^{i\vec k\vec r}d^3\vec r 
  =\frac{1}{(2\pi)^{\frac32}V^{\frac12}}\sum\limits_ie^{i\vec k\vec r_i}\int \limits_{V_i} \frac{\rho(\vec r-{\vec r}_i)}{\overline{\rho}} e^{i\vec k(\vec r-{\vec r}_i)}d^3(\vec r-{\vec r}_i) \\
  &&=\frac{1}{(2\pi)^{\frac32}\overline{\rho}V^{\frac12}}\sum\limits_ie^{i\vec k\vec r_i}m_i
  \left\{\frac{1}{m_i}\int \limits_{V_i}\rho(\vec r-{\vec r}_i) e^{i\vec k(\vec r-{\vec r}_i)}d^3(\vec r-{\vec r}_i)\right\}
  = \frac{1}{(2\pi)^{\frac32}\overline{\rho}V^{\frac12}}\sum\limits_i e^{i\vec k{\vec r}_i}m_iy_i(k) \\
  &&=\frac{1}{\overline{\rho}}\sum\limits_jn(m_j)m_j
  \left\{\frac{1}{(2\pi)^{\frac32}V^{\frac12}n(m_j)}\sum\limits_{l=1}^{N_j} e^{i\vec k{\vec r}_l}y_l(k,m_j)\right\} 
  =\frac{1}{\overline{\rho}}\int\limits_0^{\infty} m\cdot n(m)\delta_{\vec k}(m)\overline{y}(m,k)dm
\end{eqnarray*}
\end{widetext}

Here, the integration over the whole volume $V$ has been split into
integrations over volumes $V_i$, each occupied by spatially separated
halos; the Fourier transform of $i$-th density profile we denote as
$y_i(k)$, it is normalized by its masses $m_i$. The halos are binned
into the subsets with equal masses $m_j$, the number of halos is
denoted by $N_j$. Also, it was assumed that the halos of equal masses
have identical density profiles and, correspondingly, their
Fourier transforms, $\overline{y}(m,k)$, are identical too.  The
$\delta_{\vec k}(m)$ is denotation of Fourier amplitude of spatial
distribution of halos with masses $m$. The summation has been changed
to the integration.

\section[B]{Fourier modes of galaxy number density inhomogeneities}
\label{app_b}

The Fourier amplitude for relative fluctuations of galaxy
concentration takes the following form:
\begin{widetext}
\begin{eqnarray*}
  &&\delta_{g|\vec k}= \frac{1}{(2\pi)^{\frac32}V^{\frac12}}\int \limits_{V} \delta_g(\vec r) e^{i\vec k\vec r}d^3\vec r \\
&&=\frac{1}{(2\pi)^{\frac32}V^{\frac12}}\sum\limits_ie^{i\vec k\vec r_i}\int \limits_{V_i} \frac{n_g(\vec r-{\vec r}_i)}{\overline{n}_g} 
e^{i\vec k(\vec r-{\vec r}_i)}d^3(\vec r-{\vec r}_i)=\frac{1}{(2\pi)^{\frac32}\overline{n}_gV^{\frac12}}\sum\limits_ie^{i\vec k\vec r_i}N_{g|i}
  \left\{\frac{1}{N_{g|i}}\int \limits_{V_i}n_g(\vec r-{\vec r}_i) e^{i\vec k(\vec r-{\vec r}_i)}d^3(\vec r-{\vec r}_i)\right\} \\
  &&=\frac{1}{(2\pi)^{\frac32}\overline{n}_gV^{\frac12}}\sum\limits_i e^{i\vec k{\vec r}_i}N_{g|i}y_{g|i}(k)
  =\frac{1}{(2\pi)^{\frac32}\overline{n}_gV^{\frac12}}\sum\limits_m\sum\limits_{j_m} e^{i\vec k{\vec r}_{j_m}}N_{g|j_m}y_{g|j_m}(k,m) \\
  &&=\frac{1}{\overline{n}_g}\sum\limits_m \overline{y}_g(k,m)\left\{\frac{1}{(2\pi)^{\frac32}V^{\frac12}}\sum_{j_m} N_{g|j_m}e^{i\vec k{\vec r}_{j_m}}\right\} 
  =\frac{1}{\overline{n}_g}\sum\limits_m \overline{y}_g(k,m)\sum\limits_{N=0}^{\infty} N n(m,N)
  \left\{\frac{1}{(2\pi)^{\frac32}V^{\frac12}n(m,N)}\sum_{l_{mN}} e^{i\vec k{\vec r}_{l_{mN}}}\right\} \\
  &&=\frac{1}{\overline{n}_g}\sum\limits_m \overline{y}_g(k,m)\sum\limits_{N=0}^{\infty} Nn(m)p(N|m)\delta_{\vec k}(m,N)
  =\frac{1}{\overline{n}_g}\sum\limits_m n(m)\delta_{\vec k}(m)\overline{y}_g(k,m)\sum\limits_{N=0}^{\infty} Np(N|m)\\
&&=\frac{1}{\overline{n}_g}\int\limits_0^{\infty} \left<N|m\right> n(m)\delta_{\vec k}(m)\overline{y}_g(k,m)dm, 
\end{eqnarray*}
\end{widetext}

Here, as in Appendix \ref{app_a}, the integration over whole volume
$V$ was replaced by integration over the number of volumes $V_i$,
filled by spatially separated halos. The Fourier images of profiles of
concentration of galaxies in halo are normalized by their number
$N_{g|i}$; the Fourier image of $i$-th profile of galaxy concentration
is denoted by $y_{g|i}(k)$. The halo was partitioned into subsets of
equal masses $m$ and normalized by means of index $j_m$.

It was assumed that halo of equal masses have identical profiles of
galaxy concentration, so their Fourier images $\overline{y}_g(m,k)$
are identical; the sets of halos with equal masses $m$ were
partitioned into subsets containing the same number $N$ of galaxies
and denoted as $n(m,N)$. The halo concentration $n(m,N)$ is
represented as a product of concentration of all halos with mass $m$
and conditional probability of event that these halos contain $N$
galaxies each, $n(m,N) = n(m)p(N|m)$. The designation was introduced
for Fourier amplitude of spatial distribution of halos with masses
$m$ and containing $N$ galaxies as $\delta_{\vec k}(m,N)$. It was
assumed that spatial distribution of halos of mass $m$ and number of
galaxies $N$ match the spatial distribution of all halos with masses
$m$, $\delta_{\vec k}(m) = \delta_{\vec k}(m,N)$. The average number
of galaxies in halo of mass $m$ is denoted as $\left<N|m\right> =
\sum\limits_{N=0}^{\infty} Np(N|m)$, also the sum was replaced by
integration.

\label{lastpage}


\begin{thebibliography}{99}

 
\bibitem {b71} S.D.M. White, M. Rees, Mon. Not. R. Astron. Soc. \textbf{183}, 341 (1978).
\bibitem {b6} A. Benson, Phys. Rep. \textbf{495}, 33 (2010).
\bibitem {b4} C. Safranek-Shrader, V. Bromm, M. Milosavljevic, Astrophys. J. \textbf{723}, 1568 (2010).
\bibitem {b57} N. Katz, D.H. Weinberg, L. Hernquist, Astrophys. J. Suppl. \textbf{105}, 19 (1996).
\bibitem {b58} V. Springel, L. Hernquist, Mon. Not. R. Astron. Soc. \textbf{339}, 289 (2003).
\bibitem {b59} G. Kauffmann, S.D.M. White, B. Guiderdoni, Mon. Not. R. Astron. Soc. \textbf{264}, 201 (1993).
\bibitem {b60} S. Cole, A. Aragon-Salamanca, C.S. Frenk, J.F. Navarro, S.E. Zepf, Mon. Not. R. Astron. Soc. \textbf{271}, 781 (1994).
\bibitem {b61} R.S. Somerville, J.R. Primack, Mon. Not. R. Astron. Soc. \textbf{310}, 1087 (1999).
\bibitem {b62} D.J. Croton {\it et al.}, Mon. Not. R. Astron. Soc. \textbf{365}, 11 (2006).
\bibitem {b63} R.G. Bower, A.J. Benson, R. Malbon, J.C. Helly {\it et al.}, Mon. Not. R. Astron. Soc. \textbf{370}, 645 (2006).
\bibitem {b19} X. Yang, H. Mo, F. van den Bosch, Mon. Not. R. Astron. Soc. \textbf{339}, 1057 (2003).
\bibitem {b32} X. Yang, H.J. Mo, Y.P. Jing, F.C. van den Bosch, Y. Chu, Mon. Not. R. Astron. Soc. \textbf{350}, 1153 (2004).
\bibitem {b33} F.C. van den Bosch, X. Yang, H.J. Mo, S.M. Weinmann, A.V. Macci\`o et al., Mon. Not. R. Astron. Soc. \textbf{376}, 841 (2007).
\bibitem {b20} B. Moster, R. Somerville, C. Maulbetsch, F. van den Bosch et al, Astroph. J. \textbf{710}, 903 (2010).
\bibitem {r20} C. Conroy, R.H. Wechsler, A.V. Kravtsov, Astrophys. J. \textbf{647}, 201 (2006).
\bibitem {b1} Q. Guo, S. White, C. Li, M. Boylan-Kolchin, Mon. Not. R. Astron. Soc. \textbf{404}, 1111 (2010).
\bibitem {b7} J.A. Peacock, R.E. Smith, Mon. Not. R. Astron. Soc. \textbf{318}, 1144 (2000).
\bibitem {b70} C.-P. Ma, J.N. Fry, Astrophys. J. \textbf{531}, L87 (2000).
\bibitem {b65} U. Seljak, Mon. Not. R. Astron. Soc. \textbf{318}, 203 (2000).
\bibitem {b67} A.A. Berlind, D.H. Weinberg, Astrophys. J. \textbf{550}, 212 (2002).
\bibitem {r03} R. Scoccimarro, R.K. Sheth, L. Hui, B. Jain,  Astroph. J. \textbf{546}, 20 (2001).  
\bibitem {b0} A. Cooray, R. Sheth, Physics Reports, \textbf{372}, 1 (2002).
\bibitem {b29} C. Giocoli, M. Bartelmann, R.K. Sheth, M. Cacciato, Mon. Not. R. Astron. Soc. \textbf{408}, 300 (2010).
\bibitem {r04} R.K. Sheth, B. Jain, Mon. Not. R. Astron. Soc. \textbf{345}, 529 (2003).  
\bibitem {r05} R.E. Smith, P.I.R. Watts, Mon. Not. R. Astron. Soc. \textbf{360}, 203 (2005).  
\bibitem {r06} R.E. Smith, P.I.R. Watts, R.K. Sheth, Mon. Not. R. Astron. Soc. \textbf{365}, 214 (2006).   
\bibitem {b26} A.J.S. Hamilton, M. Tegmark, Mon. Not. R. Astron. Soc. \textbf{330}, 506 (2002).
\bibitem {b27} M. Tegmark, M.R. Blanton, M.A. Strauss, F. Hoyle {\it et al.}, Astrophys. J. \textbf{606}, 702 (2004).
\bibitem {b78} R.C. Tolman, {\it Relativity, thermodynamics and cosmology} (Oxford, Clearendon Press, 1969).
\bibitem {b17} Yu. Kulinich, Kinematics and Physics of Celestial Bodies \textbf{24}, 121 (2007).
\bibitem {b17a} Yu. Kulinich, B. Novosyadlyj, V. Pelykh, J. Phys. Stud. \textbf{11}, 473 (2007).
\bibitem {b18} Yu. Kulinich, B. Novosyadlyj, J. Phys. Stud., \textbf{7}, 234 (2003).
\bibitem {b12} P.J.E. Peebles, {\it The large-scale structure of the universe} (Princeton University Press, Princeton, N.J., 1980).
\bibitem {b83} V.R. Eke, S. Cole, C.S. Frenk, Mon. Not. R. Astron. Soc. \textbf{282}, 263 (1996).
\bibitem {b73} G.M. Voit, Rev. of Mod. Phys. \textbf{77}, 207 (2005).
\bibitem {b16} J. Navarro, C. Frenk, S. White, Astroph. J. \textbf{490}, 493 (1997).
\bibitem {b84} Y.P. Jing, Astroph. J. \textbf{535}, 30 (2000).
\bibitem {b86} J.S. Bullock, T.S. Kolatt, Y. Sigad, R.S. Somerville {\it et al.}, Mon. Not. R. Astron. Soc. \textbf{321}, 559 (2001).
\bibitem {b5} V. Eke, J. Navarro, M. Steinmetz, Astroph. J. \textbf{554}, 114 (2001).
\bibitem {b3} A. Duffy, J. Schaye, S. Kay, V. Dalla, Mon. Not. R. Astron. Soc. Letters \textbf{390}, L64 (2008).
\bibitem {b88} E. Komatsu, J. Dunkley, M. Nolta, C. Bennett et al, Astroph. J. Suppl., \textbf{180}, 330 (2009).
\bibitem {b34} W.H. Press, P. Schechter, Astroph. J. \textbf{187}, 425 (1974).
\bibitem {b10} J. Bond, S. Cole, G. Efstathiou, N. Kaiser, Astrophys. J. \textbf{379}, 440 (1991).
\bibitem {b36} R.G. Bower, Mon. Not. R. Astron. Soc. \textbf{248}, 332 (1991).
\bibitem {b37} C. Lacey, S. Cole, Mon. Not. R. Astron. Soc. \textbf{262}, 627 (1993).
\bibitem {b38} R.K. Sheth, G. Tormen, Mon. Not. R. Astron. Soc. \textbf{308}, 119 (1999).
\bibitem {b8} R. Sheth, H. Mo, G. Tormen, Mon. Not. R. Astron. Soc. \textbf{323}, 1 (2001).
\bibitem {b44} W. Hu, A.V. Kravtsov, Astroph. J. \textbf{584}, 702 (2003).
\bibitem {b39} M. Davis, G. Efstathiou, C.S. Frenk, S.D.M. White, Astrophys. J. \textbf{292}, 371 (1985).
\bibitem {b40} C. Lacey, S. Cole, Mon. Not. R. Astron. Soc. \textbf{271}, 676 (1994).
\bibitem {b41} A. Jenkins, C.S. Frenk, S.D.M. White, J.M. Colberg {\it et al.}, Mon. Not. R. Astron. Soc. \textbf{321}, 372 (2001).
\bibitem {b42} A.F. Neto, L. Gao, P. Bett, S. Cole {\it et al.}, Mon. Not. R. Astron. Soc. \textbf{381}, 1450 (2007).
\bibitem {b2} A. Maccio, A. Dutton, F. van den Bosch, Mon. Not. R. Astron. Soc. \textbf{391}, 1940 (2008).
\bibitem {b46} G. Kauffmann, J.M. Colberg, A. Diaferio, S.D.M. White, Mon. Not. R. Astron. Soc. \textbf{303}, 188 (1999).
\bibitem {Peebles1993} P.J.E. Peebles, {\it Principles of Physical Cosmology} (Princeton University Press, Princeton, N.J., 1993).
\bibitem {b49} G. Efstathiou, J.R. Bond, S.D.M. White, Mon. Not. R. Astron. Soc. \textbf{258}, 1 (1992).
\bibitem {b15} J. Peacock, S. Dodds, Mon. Not. R. Astron. Soc. \textbf{280}, L19 (1996).
\bibitem {b14} R. Smith, J. Peacock, A. Jenkis, S.D. White, C.S. Frenk et al, Mon. Not. R. Astron. Soc. \textbf{341}, 1311 (2003).
\bibitem {Hamilton1991} A.J.S. Hamilton, P. Kumar, E. Lu, A. Matthews, Astroph. J. \textbf{374}, L1 (1991).
\bibitem {b100} M. Viel, K. Markovi\'c, M. Baldi, J. Weller, 2011, Mon. Not. R. Astron. Soc. \textbf{421}, 50 (2012). 
\bibitem {Hansen2002} S.H. Hansen, J. Lesgourgues, S. Pastor, J. Silk, Mon. Not. R. Astron. Soc. \textbf{333}, 544 (2002).
\bibitem {Viel2005} M. Viel, J. Lesgourgues, M.G. Haehnelt, S. Matarrese, A. Riotto, Phys. Rev. D \textbf{71}, id. 063534 (2005)
\bibitem {Smith2011} R.E. Smith, K. Markovi\'c, Phys. Rev. D \textbf{84}, id. 063507 (2011).
\bibitem {b102} Q. Guo, S. Cole, V. Eke, C. Frenk, Mon. Not. R. Astron. Soc. \textbf{427}, 428 (2012).
\bibitem {camb} A. Lewis, A. Challinor and A. Lasenby, Astrophys. J. \textbf{538}, 473 (2000).
\bibitem {camb_source} http://camb.info
\bibitem {r01} F. Bernardeau, A\&A \textbf{291}, 697 (1994).
\bibitem {r02} R.K. Sheth,  Mon. Not. R. Astron. Soc. \textbf{300}, 1057 (1998). 
\bibitem {r07} D.H. Zhao, Y.P. Jing, H.J. Mo, G. Borner, Astroph. J. \textbf{707}, 354 (2009).  
\bibitem {r08} A. Schneider, R.E. Smith, A.V. Maccio, B. Moore, Mon. Not. R. Astron. Soc. \textbf{424}, 684 (2012).
\bibitem {r10} P. Valageas, T. Nishimichi, Astronomy \& Astrophysics, \textbf{527}, id.A87 (2011).  
\bibitem {r11} R.E. Smith, K. Markovic, Physical Review D, \textbf{84}, id. 063507 (2011).
\bibitem {r12} P. Bode, J.P. Ostriker, N. Turok, Astroph. J., \textbf{556}, 93, (2001).
\end{thebibliography}
\end{document}